\documentclass[11pt]{article}
\pdfoutput=1
\usepackage{jheppub}
\usepackage{physics}
\usepackage{bbm}
\usepackage{empheq}
\usepackage{tikz}
\usepackage{cancel}
\usepackage{amsthm}
\usepackage{verbatim}

\newtheorem*{theorem*}{Theorem}
\theoremstyle{definition}

\newtheorem*{definition*}{Definition}

\newcommand\nn{\nonumber}

\newcommand\fft[2]{\frac{#1}{#2}}

\newcommand\ri{\mathfrak{i}}

\newcommand\bse{\begin{subequations}}
\newcommand\ese{\end{subequations}}

\newcommand\mF{\mathcal{F}}
\newcommand\mN{\mathcal{N}}
\newcommand\mM{\mathcal{M}}
\newcommand\mX{\mathcal{X}}

\newcommand\tF{\widetilde{F}}

\newcommand\Vol{\text{Vol}}
\newcommand\vol{\text{vol}}

\makeatletter
\newcommand*{\rom}[1]{\expandafter\@slowromancap\romannumeral #1@}
\makeatother

\begin{document}

\preprint{CQUeST-2026-0778}
\title{Type IIB Supergravity Action and Holography}

\author[a]{Soumya Adhikari,}
\author[a]{Junho Hong,}
\author[a]{Chanyoung Joung,}
\author[a]{and Geum Lee}
\affiliation[a]{Department of Physics \& Center for Quantum Spacetime, Sogang University\,,\\ 35 Baekbeom-ro, Mapo-gu, Seoul 04107, Republic of Korea}

\emailAdd{ssoumya.a012@gmail.com}
\emailAdd{junhohong@sogang.ac.kr}
\emailAdd{chanyoung.joung@sogang.ac.kr}
\emailAdd{mbong1239@gmail.com}
	
\abstract{In the prototypical AdS$_5$/CFT$_4$ correspondence, the free energy of $\mN=4$ SU$(N)$ super Yang-Mills theory is commonly reproduced from the Euclidean on-shell action of five-dimensional gauged supergravity -- a consistent truncation of Type IIB supergravity -- rather than computed directly in ten dimensions. A longstanding obstacle to the latter is that the conventional Type IIB pseudo-action evaluated on the $AdS_5\times S^5$ background vanishes identically, apparently precluding a first-principles holographic comparison. A recent proposal by Kurlyand and Tseytlin, based on the Pasti-Sorokin-Tonin formulation, resolves this issue for a special class of backgrounds including the $AdS_5\times S^5$ vacuum by introducing a topological term required for consistency, yielding a non-vanishing on-shell value in agreement with holography. In this work we extend this refinement to a broader class of Type IIB backgrounds by introducing a generalized topological correction under milder conditions, encompassing AdS geometries of generic dimension and non-vanishing 2-form potentials. We test the proposal on non-trivial solutions such as the Lunin–Maldacena background and the $AdS_4$ $S$-fold solution, and find agreement with the corresponding lower-dimensional gauged supergravity on-shell actions and thereby with the expected holographic observables. Our results place direct holographic comparisons within the ten-dimensional Type IIB framework on firmer ground.}
	
\maketitle \flushbottom

\allowdisplaybreaks

\section{Introduction}\label{Introduction}
The AdS/CFT correspondence traditionally identifies the Euclidean gravitational path integral evaluated around an AdS background with the partition function of a dual CFT \cite{Maldacena:1997re,Witten:1998qj,Gubser:1998bc}. In the prototypical duality between Type IIB superstring theory on $AdS_5\times S^5$ and $\mN=4$ $\mathrm{SU}(N)$ super Yang–Mills (SYM) theory, such a comparison leads to a precise agreement in the double scaling limit -- the 't~Hooft limit followed by a strong coupling limit: the regulated Euclidean on-shell action of 5d gauged supergravity reproduces the planar free energy of the dual SYM theory, whose universal logarithmic divergence is governed by the conformal anomaly \cite{Henningson:1998gx}. Moreover, supersymmetric localization \cite{Pestun:2007rz} has enabled us to probe this holographic correspondence beyond the semi-classical regime, providing a powerful tool for testing holography as well as exploring a wide range of questions in quantum gravity and strongly coupled CFTs.

However, the proper holographic counterpart of the $\mN=4$ SYM free energy in the double scaling limit is the Type IIB supergravity action evaluated on the $AdS_5\times S^5$ background, rather than the 5d gauged supergravity on-shell action. Although 5d gauged supergravity arises from a consistent truncation of Type IIB supergravity \cite{Cvetic:1999xp,Cvetic:2000nc,Baguet:2015sma}, the absence of scale separation between the AdS geometry and the internal space implies that this lower-dimensional theory cannot be regarded as a complete effective field theory for the full string dynamics. This limitation has been demonstrated explicitly in the closely related context of AdS$_4$/CFT$_3$: an 11d supergravity 1-loop computation reproduces the logarithmic term in the large $N$ expansion of the ABJM free energy \cite{Bhattacharyya:2012ye}, whereas 4d gauged supergravity fails to do so unless the contributions from the entire Kaluza-Klein tower on the internal manifold are included \cite{Bobev:2023dwx}. This illustrates that lower-dimensional gauged supergravity captures only a restricted subsector of the bulk quantum dynamics, and that a first-principles holographic comparison should therefore be formulated directly in the 10d or 11d parent theory.

A direct holographic comparison based on Type IIB supergravity, however, encounters a well-known subtlety. The Type IIB supergravity pseudo-action vanishes on shell for the $AdS_5\times S^5$ background, and therefore does not reproduce the planar free energy of the dual $\mN=4$ SYM theory, in contrast with the non-vanishing result obtained from the lower-dimensional gauged supergravity on-shell action. From a purely classical supergravity perspective, such a mismatch between on-shell actions across dimensions does not necessarily signal an inconsistency, since the higher-dimensional theory and its consistent truncation remain fully compatible at the level of equations of motion, despite differences at the level of action \cite{Cvetic:1999xp}. When analyzing a gravitational path integral through a saddle point approximation, however, the on-shell action governs the leading semi-classical contribution to the path integral and therefore discrepancies at the level of the action cannot simply be regarded as innocuous. In the context of the AdS$_5$/CFT$_4$ correspondence of interest, this further implies that the vanishing of the Type IIB supergravity on-shell action on the $AdS_5\times S^5$ background must be properly remedied, as it appears to be in tension with the non-vanishing free energy of the dual field theory; the successful matching obtained using the lower-dimensional gauged supergravity action cannot by itself resolve this issue. 

A possible resolution to this problem was recently proposed in \cite{Kurlyand:2022vzv}. The starting point is the PST formulation of Type IIB supergravity \cite{DallAgata:1997gnw,DallAgata:1998ahf}, which modifies the usual pseudo-action by introducing an auxiliary scalar field in a non-polynomial fashion, following the general framework of \cite{Pasti:1995ii,Pasti:1995tn,Pasti:1996vs} for constructing covariant actions for chiral $p$-forms. The resulting PST action possesses additional gauge symmetries, and its equations of motion, along with an appropriate gauge fixing, reproduce the self-duality constraint on the 5-form field strength as well as the standard Type IIB equations of motion. In other words, the PST framework improves the pseudo-action formulation where the self-duality condition must be imposed by hand at the level of the equations of motion, by incorporating the self-duality constraint dynamically within the action principle. The key observation of \cite{Kurlyand:2022vzv} is that, when applied to Type IIB backgrounds with boundaries, including the case of an AdS geometry regulated by a radial cutoff, the PST action requires an additional topological term. When evaluated on the $AdS_5\times S^5$ solution, this additional contribution yields a non-vanishing on-shell action that precisely matches the result obtained from the 5d gauged supergravity description and therefore reproduces the planar free energy of the dual $\mN=4$ $\mathrm{SU}(N)$ SYM theory. Importantly, the appearance of this topological correction is not introduced ad hoc to enforce holographic matching; rather, it arises from an independent requirement -- the dynamical derivation of the self-duality for the 5-form field strength within the PST framework. Related topological modifications of Type IIB supergravity have subsequently been discussed from different perspectives in \cite{Mkrtchyan:2022xrm,Chakrabarti:2022jcb,Frey:2024tnn}.

An important limitation of the refined PST formulation proposed in \cite{Kurlyand:2022vzv} is that its derivation relies on a set of restrictive assumptions, which exclude a variety of holographically relevant AdS backgrounds beyond the simplest $AdS_5\times S^5$ vacuum of Type IIB supergravity. In particular, the justification of the topological correction given in \cite{Kurlyand:2022vzv} applies to backgrounds where both the external and internal manifolds have dimension five and the NSNS and RR 2-form fields vanish. However, many well-studied holographic backgrounds lie outside this restricted class and nevertheless admit well-established dual field theories. Therefore, if the refined PST formulation is to provide a general resolution of the discrepancy between the Type IIB on-shell action and the free energy of the dual field theory, it is essential to extend the refinement proposed in \cite{Kurlyand:2022vzv} to a broader class of Type IIB AdS backgrounds.

In this paper we take a first step in this direction by refining the PST formulation of Type IIB supergravity through the addition of an appropriate topological term, in the spirit of \cite{Kurlyand:2022vzv}, but now for a substantially larger class of backgrounds. In particular, our proposal applies to configurations in which the AdS factor is not necessarily five-dimensional and the NSNS and RR 2-form fields may be non-vanishing. As a non-trivial test of our proposal, we evaluate the resulting PST action on two well-known Type IIB backgrounds studied extensively in the context of holographic duality: the Lunin–Maldacena solution dual to the $\mN=1$ exactly marginal deformation of $\mN=4$ SYM theory \cite{Lunin:2005jy,Leigh:1995ep} and the $AdS_4\times S^1\times S^5_\text{deformed}$ $S$-fold background \cite{Inverso:2016eet} dual to a 3d $\mN=4$ $S$-fold CFT addressed in \cite{Assel:2018vtq}. In both cases, we find that our refined PST-IIB action produces the expected non-vanishing on-shell value, in agreement with the dual field-theory observable. This provides further support that a properly refined Type IIB action enables a direct, first-principles holographic comparison between the supergravity on-shell action and the free energy of the dual CFT, without relying on the lower-dimensional gauged supergravity truncation.

This paper is organized as follows. In Section \ref{sec:IIB}, we review Euclidean Type IIB supergravity, focusing on the PST formulation in which the self-duality of the 5-form field strength arises dynamically from the action principle, and briefly comment on alternative formulations including the string-field-theory-motivated and clone-field approaches. In Section \ref{sec:proposal}, we propose a refined PST formulation of Euclidean Type IIB supergravity by introducing a suitable topological term that restores the relevant gauge symmetry in the presence of boundaries under significantly weaker assumptions than those employed in \cite{Kurlyand:2022vzv}. We also clarify the relation between this refined PST formulation and the clone-field formalism, providing a prescription that resolves the ambiguity in the decomposition of the self-dual 5-form field strength. In Section \ref{sec:example}, we test our proposal by evaluating the refined PST-IIB action on several holographically relevant backgrounds -- the Lunin–Maldacena geometry and the $AdS_4$ $S$-fold solution, and demonstrate agreement with the corresponding lower-dimensional gauged supergravity results and the expected dual field-theory observables. We conclude in Section \ref{sec:discussion} with a summary of our results and possible future directions.

\section{Euclidean Type IIB Supergravity}\label{sec:IIB}
This section reviews Euclidean Type IIB supergravity focusing on the bosonic part, which underlies the holographic comparisons performed in this work. We begin in Subsection \ref{sec:pse} by establishing our conventions in Euclidean signature by Wick rotating the standard Lorentzian bosonic pseudo-action of Type IIB supergravity, where the self-duality constraint on the 5-form field strength is imposed by hand. In Subsection \ref{sec:IIB:PST}, we revisit the Pasti-Sorokin-Tonin (PST) formulation \cite{Pasti:1995ii,Pasti:1995tn,Pasti:1996vs} applied to Type IIB supergravity \cite{DallAgata:1997gnw,DallAgata:1998ahf} in Euclidean signature following the conventions in the previous subsection, where the self-duality of the 5-form field strength emerges dynamically as a consequence of the equations of motion. Finally, in Subsection \ref{sec:IIB:other}, we briefly review alternative formulations of the Type IIB supergravity action and comment on the approach that maintains manifest duality symmetry \cite{Mkrtchyan:2022xrm,Evnin:2023ypu}.

\subsection{Pseudo Action for Euclidean Type IIB Supergravity}\label{sec:pse}
The bosonic field content of Lorentzian Type IIB supergravity consists of the NS-NS sector $(g^\text{L}_{\mu\nu},B_2,\phi)$ and the R-R sector $(C_0,C_2,C_4)$ with associated field strengths defined by
\begin{align}
	H_3&=dB_2\,,\quad F_{n+1}=dC_n\,,\quad \tF_3=F_3-C_0H_3\,, \nn\\
    \tF_5&=F_5-\fft12C_2\wedge H_3+\fft12B_2\wedge F_3=F_5+\mX_5\,.\label{IIB:content}
\end{align}
The bosonic sector of Lorentzian Type IIB supergravity is described by a `pseudo-action', which in the Einstein frame takes the form
\begin{align}
	S_\text{IIB,L}&=\fft{1}{2\kappa^2}\int \bigg(*_\text{L}R-\fft12d\phi\wedge*_\text{L}d\phi-\fft12e^{-\phi}H_3\wedge*_\text{L}H_3\bigg)\nn\\
	&\quad-\fft{1}{4\kappa^2}\int \bigg(e^{2\phi}F_1\wedge*_\text{L}F_1+e^{\phi}\tF_3\wedge*_\text{L}\tF_3+\fft12\tF_5\wedge*_\text{L}\tF_5\bigg)\nn\\
	&\quad-\fft{1}{4\kappa^2}\int C_4\wedge H_3\wedge F_3\,.\label{IIB:action:L}
\end{align}
We largely follow the standard conventions in the literature, see \cite{Becker:2006dvp,Hamilton:2016ito} for example, while our notation for differential forms and related conventions is summarized in Appendix~\ref{app:convention:diff}. Superscripts
and subscripts ``L'' are used to indicate Lorentzian signature for clarity.

The action \eqref{IIB:action:L} is commonly referred to as a `pseudo-action', since it does not yield the self-duality constraint on the 5-form field strength $\tF_5$,
\begin{equation}
	*_{\mathrm L}\tF_5=\tF_5 \,,
\end{equation}
as a consequence of the equations of motion. Instead, this condition must be imposed separately at the level of the field equations in order to recover the correct dynamics of Type IIB supergravity \cite{Bergshoeff:1995sq}.

\medskip

We now implement the Wick rotation following the conventions summarized in Appendix~\ref{app:convention:Wick}, thereby obtaining the Euclidean counterpart of the Lorentzian Type IIB supergravity reviewed above. The resulting pseudo-action for the bosonic sector of Euclidean Type IIB supergravity reads ($S=-\ri S_\text{L}$)\footnote{The Chern-Simons (CS) term may differ from other presentations in the literature by an overall factor of $i$, reflecting a choice of conventions in the Wick rotation.}
\begin{align}
	S_\text{IIB}&=-\fft{1}{2\kappa^2}\int \bigg(*R-\fft12d\phi\wedge*d\phi-\fft12e^{-\phi}H_3\wedge*H_3\bigg)\nn\\
	&\quad+\fft{1}{4\kappa^2}\int \bigg(e^{2\phi}F_1\wedge*F_1+e^{\phi}\tF_3\wedge*\tF_3+\fft12\tF_5\wedge*\tF_5\bigg)\nn\\
	&\quad+\fft{i}{4\kappa^2}\int C_4\wedge H_3\wedge F_3\,.\label{IIB:action}
\end{align}
For notational simplicity, we henceforth omit explicit superscripts or subscripts distinguishing Euclidean quantities. As in the Lorentzian case, the Euclidean action \eqref{IIB:action} remains a pseudo-action. A subtle difference is that, due to the choice of conventions, the 5-form field strength obeys the self-duality condition
\begin{align}
	*\tF_5=i\tF_5 \label{self-dual}
\end{align}
in Euclidean signature, which must again be imposed by hand at the level of the equations of motion.

\medskip

The equations of motion obtained by varying the Euclidean Type IIB pseudo-action \eqref{IIB:action} read\footnote{$|F_p|^2=\frac{1}{p!}F_{\mu_1\cdots\mu_p}F^{\mu_1\cdots\mu_p}$} 
\begin{subequations}
\begin{align}
    \delta_{g_{\mu\nu}}:\quad0&= R_{\mu\nu}-\frac{1}{2}\partial_\mu\phi\partial_\nu\phi-\frac{1}{2}e^{2\phi}F_{\mu} F_{\nu}-\frac{1}{96}(\tF_{\mu\mu_1\mu_2\mu_3\mu_4}\tF_{\nu}{}^{\mu_1\mu_2\mu_3\mu_4})\nonumber\\ &\quad-\frac{1}{4}e^{\phi}(\tF_{\mu\mu_1\mu_2}\tF_{\nu}{}^{\mu_1\mu_2}-\frac{1}{2}g_{\mu\nu}|\tF_3|^2)-\frac{1}{4}e^{-\phi}(H_{\mu\mu_1\mu_2}H_{\nu}{}^{\mu_1\mu_2}-\frac{1}{2}g_{\mu\nu}|H_3|^2)\,,\\
    \delta_{B_2}:\quad0&=d(e^{-\phi}*H_3)-i\tF_3\wedge\tF_5-e^{\phi}F_1\wedge*\tF_3\,,\\
    \delta_{\phi}:\quad0&=d*d\phi +\frac{1}{2}e^{-\phi} H_3\wedge*H_3-e^{2\phi}F_1\wedge*F_1-\frac{1}{2}e^{\phi}\tF_3\wedge*\tF_3\,,\\
    \delta_{C_0}:\quad0&=d(e^{2\phi}*F_1)+e^{\phi}H_3\wedge*\tF_3\,,\\
    \delta_{C_2}:\quad0&=d(e^\phi*\tF_3)+i H_3\wedge\tF_5\,,\\
    \delta_{C_4}:\quad0&=d*\tF_5-i H_3\wedge F_3\,\label{IIB:eom-C4}.
\end{align}\label{IIB:eom}%
\end{subequations}
These equations of motion, together with the self-duality constraint \eqref{self-dual}, constitute the dynamics of the bosonic sector of Euclidean Type IIB supergravity. In the following subsection, we will demonstrate how the PST formulation naturally reproduces this set of equations from a manifestly gauge-invariant action principle, without the need to impose the duality constraint by hand.

\subsection{PST Formulation of Euclidean Type IIB supergravity}\label{sec:IIB:PST}
We aim to construct a bosonic action of Euclidean Type IIB supergravity from which the self-duality constraint \eqref{self-dual} can be derived as a consequence of the equations of motion. The PST formulation provides such an action in Lorentzian signature \cite{DallAgata:1997gnw,DallAgata:1998ahf}, improving on the pseudo-action \eqref{IIB:action:L}. Its Euclidean counterpart may be formulated in an analogous manner by introducing an auxiliary scalar field $a(x)$ in the pseudo-action \eqref{IIB:action} as \cite{Kurlyand:2022vzv}
\begin{align}
    S_\text{PST}=S_\text{IIB}+\fft{1}{8\kappa^2}\int\ri_v\mF_5\wedge*\ri_v\mF_5\,,\label{PST-IIB:action}
\end{align}
where we have also introduced an anti-self-dual 5-form field $\mF_5$ and a vector field $v=v^\mu\partial_\mu$ associated with the auxiliary scalar field as
\begin{align}
    \mF_5\equiv \tF_5+i *\tF_5\qquad\&\qquad v_\mu\equiv\fft{1}{\sqrt{-|
    \partial a|^2}}\partial_\mu a\,.\label{calF5}
\end{align}
The essential modification relative to the pseudo-action is the additional term involving $\ri_v \mF_5$, which, as we will demonstrate, enforces the self-duality condition through
the equations of motion.

\medskip

To proceed, we observe that the PST-IIB action \eqref{PST-IIB:action} has gauge symmetries that are absent in the original pseudo-action \eqref{IIB:action}. Specifically, the action is invariant under the following two gauge transformations, 
\begin{subequations}
\begin{align}
    \delta_\eta:&\quad\delta_\eta a=\eta\,,\qquad \delta_\eta C_4=-\fft{\eta}{\sqrt{-|\partial a|^2}}\ri_v\mF_5\,,\label{PST-IIB:gg1} \\
    \delta_\ell:&\quad\delta_\ell a=0\,,\qquad \delta_\ell C_4=\ell_4\qquad(da\wedge d\ell_4=0)\,.\label{PST-IIB:gg2}
\end{align}\label{PST-IIB:gg}%
\end{subequations}
The first is equivalent to one of the original PST gauge symmetries of \cite{DallAgata:1997gnw,DallAgata:1998ahf}. The second \eqref{PST-IIB:gg2} generalizes the other
gauge symmetry presented in \cite{DallAgata:1997gnw,DallAgata:1998ahf},
as well as its extension to backgrounds with non-trivial
topology \cite{Isono:2014bsa}: rather than demanding the stronger
condition $d\ell_4 = 0$, we require only $da \wedge d\ell_4 = 0$. Note that the original PST symmetry of \cite{DallAgata:1997gnw,DallAgata:1998ahf}
is recovered as the special case $\ell_4 = \xi_3 \wedge da$. 

As verified explicitly in Appendix \ref{app:gauge-inv}, the Euclidean PST-IIB action \eqref{PST-IIB:action} is invariant under both gauge transformations \eqref{PST-IIB:gg}, provided that the gauge parameters vanish on the boundary as 
\begin{align}
    \eta|_{\partial \mM}=\ell_4|_{\partial \mM}=0\,.\label{gauge:bdry}
\end{align}
Note that this requirement is automatically satisfied for 10d backgrounds $\mM$ without boundary ($\partial\mM=\varnothing$). After introducing the PST formulation under the assumption \eqref{gauge:bdry}, we will return to how it may be improved to accommodate more general 10d backgrounds, thereby relaxing the boundary condition imposed above.

\medskip

We next derive the equations of motion by varying the Euclidean PST--IIB action \eqref{PST-IIB:action}. Compared with the original Type IIB equations of motion \eqref{IIB:eom}, the resulting field equations receive additional contributions due to the auxiliary scalar field $a(x)$ and the PST term involving $\ri_v\mF_5$. The field equations are
\begin{subequations}
\begin{align}
    \delta_{g_{\mu\nu}}:\quad0&= R_{\mu\nu}-\frac{1}{2}\partial_\mu\phi\partial_\nu\phi-\frac{1}{2}e^{2\phi}F_{\mu} F_{\nu}-\frac{1}{96}(\tF_{\mu\mu_1\mu_2\mu_3\mu_4}\tF_{\nu}{}^{\mu_1\mu_2\mu_3\mu_4}-12g_{\mu\nu}|\tF_5|^2)\nonumber\\ &\quad-\frac{1}{4}e^{\phi}(\tF_{\mu\mu_1\mu_2}\tF_{\nu}{}^{\mu_1\mu_2}-\frac{1}{2}g_{\mu\nu}|\tF_3|^2)-\frac{1}{4}e^{-\phi}(H_{\mu\mu_1\mu_2}H_{\nu}{}^{\mu_1\mu_2}-\frac{1}{2}g_{\mu\nu}|H_3|^2)\nonumber\\ &\quad-\frac{1}{24}((\ri_v\mF_5)_{\mu\mu_1\mu_2\mu_3}(\ri_v\mF_5)_{\nu}{}^{\mu_1\mu_2\mu_3}-\frac{9}{4}g_{\mu\nu}|\ri_v\mF_5|^2)\,,\label{PST-IIB:eom:g}\\
    \delta_{B_2}:\quad0&=d(e^{-\phi}*H_3)-i\tF_3\wedge\tF_5-e^{\phi}F_1\wedge*\tF_3+\frac{i}{2}C_0d(v\wedge B_2\wedge \ri_v\mF_5)-\frac{i}{2}C_0 v\wedge H_3\wedge\ri_v\mF_5\nn\\ &\quad-\frac{i}{2}d(v\wedge C_2\wedge\ri_v\mF_5)+\frac{i}{2}v\wedge F_3\wedge\ri_v\mF_5\,,\\
    \delta_{C_2}:\quad0&=d(e^{\phi}*\tF_3)+iH_3\wedge\tF_5+\frac{i}{2}d(v\wedge B_2\wedge\ri_v\mF_5)-\frac{i}{2}v\wedge H_3\wedge\ri_v\mF_5\,,\\
    \delta_{C_4}:\quad0&=d(v\wedge\ri_v\mF_5)\,,\label{PST-IIB:eom-C4}\\
    \delta_{a}:\quad0&=d(\frac{1}{\sqrt{-|\partial a|^2}}v\wedge\ri_v\mF_5\wedge\ri_v\mF_5)\label{PST-IIB:eom-a}\,,
\end{align}\label{PST-IIB:eom}%
\end{subequations}
where we have omitted the physical scalar equations of motion, which coincide with those in \eqref{IIB:eom}.

\medskip

The central claim of the PST formulation of Euclidean Type IIB supergravity may now be stated succinctly: upon an appropriate gauge fixing of the PST symmetries \eqref{PST-IIB:gg}, the equations of motion derived from the PST-IIB action \eqref{PST-IIB:eom} reproduce both the original Type IIB equations of motion \eqref{IIB:eom} and the self-duality constraint \eqref{self-dual}. Below we review this statement in detail, carefully elucidating the role of the boundary assumption \eqref{gauge:bdry}.

\begin{enumerate}
    \item By the Poincar\'e lemma, the $C_4$ equation of motion (\ref{PST-IIB:eom-C4}) implies locally the existence of a 4-form $\Phi_4$ such that
   \begin{align}
        v\wedge\ri_v\mF_5&=d\Phi_4 \qquad \text{satisfying}\qquad da\wedge d\Phi_4=0\,.\label{step1}
    \end{align}

    \item The variation of the quantity $v \wedge \ri_v \mF_5$, whose exterior derivative gives the $C_4$ equations of motion \eqref{PST-IIB:eom-C4}, under the $\delta_\ell$ gauge transformation \eqref{PST-IIB:gg2} yields 
    \begin{align}
        \delta_\ell (v \wedge \ri_v \mF_5) = v \wedge \ri_v \mF_5^{(\ell)} - v \wedge \ri_v \mF_5  = - \delta_\ell F_5=-d\ell_4 \,.
    \end{align}
    Consequently, the transformed $C_4$ equations of motion are governed by
    \begin{align}
        v \wedge \ri_v \mF_5^{(\ell)} = d(\Phi_4 - \ell_4) \,, \label{eq:Step2}
    \end{align}
    where we have used the definition of $\Phi_4$ given in the previous step \eqref{step1}.

    \item We now fix the $\delta_\ell$ gauge symmetry by choosing $\ell_4 = \Phi_4$, which implies
    \begin{align}
        v \wedge \ri_v \mF_5^{(\Phi)} = 0 \,. \label{eq:vanish}
    \end{align}
    With this choice, the transformed $C_4$ equation of motion is satisfied identically.
    
    It is important to stress, however, that this gauge-fixing procedure is valid only for backgrounds in which the 4-form $\Phi_4$ introduced in \eqref{step1} vanishes at the boundary. If $\Phi_4$ does not vanish on $\partial\mM$, the identification
	$\ell_4=\Phi_4$ is not admissible, since the $\delta_\ell$ transformation \eqref{PST-IIB:gg2} ceases to be a legitimate gauge symmetry when the gauge parameter is nonvanishing at the boundary. This observation shows that the standard PST formulation	implicitly relies on the boundary assumption \eqref{gauge:bdry}, and must therefore be supplemented by suitable boundary terms in order to treat more general backgrounds. We return to this improvement shortly.

    \item Finally, since $\mF_5$ is anti-self-dual by definition, it obeys the identity \eqref{Identity:5} as
    \begin{align}
        \mF_5^{(\Phi)} = - v \wedge \ri_v \mF_5^{(\Phi)} - i * (v \wedge \ri_v \mF_5^{(\Phi)}) \,. \label{deformed A.7g}
    \end{align}
    Substituting the gauge-fixed result \eqref{eq:vanish} then yields
    \begin{align}
        \mF_5^{(\Phi)} = 0 \,.
    \end{align}
    This vanishing condition implies that the self-duality $*\tF_5 = i \tF_5$ holds under the gauge choice made in the 3rd step. Furthermore, since all PST-specific terms appearing in the equations of motion \eqref{PST-IIB:eom} are proportional to $\mF_5$, they vanish identically under the same gauge-fixing condition. Consequently, the PST-IIB equations of motion reduce exactly to the original Type IIB supergravity equations of motion \eqref{IIB:eom}, thereby completing the proof of the claim.
\end{enumerate}
%

\medskip
\noindent\textbf{Improvement of PST formulation}
\medskip

As demonstrated in the preceding derivation, the standard PST formulation successfully reproduces the self-duality of the 5-form field strength $\tF_5$ at the level of the equations of motion. The gauge-fixing procedure, however, exposes a subtle mathematical obstruction: the dynamically determined gauge parameter $\Phi_4$ generically acquires non-vanishing boundary values, whereas the $\delta_\ell$ transformation \eqref{PST-IIB:gg2} is a valid gauge symmetry of the PST-IIB action \eqref{PST-IIB:action} only when the gauge parameter $\ell_4$ vanishes at the boundary. This incompatibility indicates that the PST-IIB action \eqref{PST-IIB:action} is incomplete in the presence of boundaries and must be augmented by appropriate boundary contributions.

Motivated by this observation, an additional boundary term supplementing the PST-IIB action has been proposed in \cite{Kurlyand:2022vzv}, leading to the modified action
\begin{align}
    S_\text{KT} = S_\text{PST} + S_\text{KT}^\text{(top)}\qquad \&\qquad S_\text{KT}^\text{(top)} = - \fft{i}{4\kappa^2}\int \tF_{5M}\wedge\tF_{5X} \,,\label{PST-IIB:action:2022}
\end{align}
which is justified for a restricted class of Type IIB backgrounds obeying the following conditions. 
\begin{itemize}
    \item The 10d spacetime topology factorizes as $M_5\times X_5$ with a non-compact $M_5$ and a compact $X_5$, while the metric is unwarped with respect to the internal $X_5$ coordinates.

    \item All IIB fields vanish except for the metric and the self-dual 5-form field strength, so that the latter admits a natural decomposition into an external component $\tF_{5M}$ and an internal component $\tF_{5X}$ as 
    \begin{align}
        \tF_5=\tF_{5M}+\tF_{5X}\qquad(*\tF_{5M}=\ri \tF_{5X})\,,\label{F5:decomposition:old}
    \end{align}
    while the additional term $S_\text{KT}^\text{(top)}$ in \eqref{PST-IIB:action:2022} becomes purely topological. In fact, \cite{Kurlyand:2022vzv} proposed a more general decomposition into `electric' and `magnetic' components motivated by holographic comparisons, which does not necessarily require the two parts to be supported exclusively on $M_5$ and $X_5$ respectively. Incorporating this generalized decomposition into a consistent refinement of the PST framework, however, involves several non-trivial points that must be carefully addressed. These will be discussed in detail in Section \ref{sec:proposal:PST}. 

    \item The original Type IIB pseudo-action \eqref{IIB:action} vanishes when evaluated on-shell.
\end{itemize}
For backgrounds satisfying these criteria -- of which the $AdS_5\times S^5$ vacuum provides a canonical example -- the improved PST-IIB action \eqref{PST-IIB:action:2022} restores the $\delta_\ell$ gauge transformation \eqref{PST-IIB:gg2} even in the presence of non-trivial boundaries. Consequently, for this restricted class of geometries, the standard PST procedure outlined in this subsection remains valid.

\medskip

The improved PST-IIB action \eqref{PST-IIB:action:2022} has been further corroborated by a completely independent line of reasoning. Specifically, the non-vanishing on-shell value of the improved PST-IIB action \eqref{PST-IIB:action:2022} for the $EAdS_5 \times S^5$ background -- entirely originating from the newly introduced topological term -- was shown to coincide precisely with the on-shell action of 5d minimal gauged supergravity evaluated on $EAdS_5$ \cite{Kurlyand:2022vzv}, as obtained from consistent truncation \cite{Cvetic:1999xp,Lu:1999bw,Cvetic:2000nc,Baguet:2015sma}. This result, in turn, matches the planar $S^4$ free energy of the dual $\mN=4$ $\text{SU}(N)$ super Yang-Mills theory. Below we highlight the physical significance of this exact agreement from three perspectives.
\begin{itemize}
    \item The prototypical AdS$_5$/CFT$_4$ correspondence under consideration is fundamentally a duality between Type IIB string theory on $AdS_5\times S^5$ and $\mN=4$ $\text{SU}(N)$ super Yang-Mills, rather than between the latter and 5d gauged supergravity. Consequently, a proper test of holographic duality should, in principle, be performed directly within the 10d Type IIB framework, even though lower-dimensional gauged supergravity often provides a convenient effective description. The improved PST-IIB action \eqref{PST-IIB:action:2022} makes such a direct holographic comparison possible at the level of on-shell actions within the original ten-dimensional setup.

    \item Prior to the introduction of the topological term in \eqref{PST-IIB:action:2022}, the Type IIB pseudo-action \eqref{IIB:action} evaluated on the $EAdS_5 \times S^5$ background vanishes identically, whereas the corresponding 5d gauged supergravity action yields a non-zero on-shell value. As a result, the notion of consistent truncation could be meaningfully formulated only at the level of equations of motion, but not at the level of the action itself \cite{Cvetic:1999xp}. While this limitation is inconsequential in purely classical analyses -- where the action serves merely as a device for deriving equations of motion -- it becomes problematic in the context of Euclidean quantum gravity, where the classical on-shell action directly governs the semi-classical approximation to the path integral and thus carries intrinsic physical significance, particularly when compared with dual field theoretic quantities. The improved PST-IIB action \eqref{PST-IIB:action:2022} resolves this tension by aligning the on-shell actions of the higher- and lower-dimensional supergravity theories related by consistent truncation.

    \item Importantly, the motivation for improving the PST-IIB action is entirely independent of holography, arising instead from the requirement that the original PST construction remain valid in the presence of boundaries. The fact that this modification simultaneously resolves a long-standing mismatch puzzle in the AdS$_5$/CFT$_4$ correspondence provides compelling evidence that the improved formulation captures the correct structure of the Type IIB action.
\end{itemize}

\medskip 

Despite this refinement, the improved PST–IIB action \eqref{PST-IIB:action:2022} remains of limited scope, and must be further generalized to accommodate the broader class of holographically relevant Type IIB backgrounds. Indeed, many AdS backgrounds of Type IIB supergravity that have been extensively studied in the holographic context fail to satisfy the assumptions underlying \eqref{PST-IIB:action:2022}. Representative examples include various $AdS_d\times X_{10-d}$ backgrounds with $d\neq 5$ and a generic internal manifold $X_{10-d}$, as well as backgrounds featuring non-vanishing NS-NS or R-R 2-form fluxes, or a non-trivial warping factor supported on the internal manifold. In Section \ref{sec:proposal}, we further extend the improved PST-IIB action \eqref{PST-IIB:action:2022} to encompass such AdS backgrounds.

\subsection{Other Formulations}\label{sec:IIB:other}
Before turning to the generalization of the improved Type IIB supergravity action \eqref{PST-IIB:action:2022} within the PST framework, we comment on alternative formulations of the same theory that address the implementation of the self-duality constraint on the 5-form field strength. In particular, we briefly introduce how a boundary term analogous to that appearing in the PST formulation as in \eqref{PST-IIB:action:2022} arises in two distinct formulations of the Type IIB supergravity action: the string field theory inspired formulation \cite{Sen:2015nph,Chakrabarti:2022jcb} and the clone-field formulation \cite{Mkrtchyan:2019opf,Mkrtchyan:2022xrm}. For a concise review of these approaches, together with the PST formulation itself, see for example \cite{Evnin:2022kqn}. A concrete derivation of the equivalence between the PST and clone-field formulations modulo boundary terms in the more general context of non-linear chiral $p$-form theories can be found in a recent work \cite{Hutomo:2025dfx}.

\medskip
\noindent\textbf{String-Field-Theory-Motivated Formulation}
\medskip

The first alternative consists in adding a suitable boundary term by hand \cite{Chakrabarti:2022jcb} to Sen's formulation of the Type IIB supergravity action \cite{Sen:2015nph,Sen:2019qit}, which is motivated from the corresponding closed string field theory action \cite{Sen:2015uaa}. It was shown that, while the off-shell structure differs from other formulations and in particular involves non-physical auxiliary fields, the proposed boundary term reduces on-shell to a purely topological contribution that reproduces the non-vanishing free energy for the $AdS_5 \times S^5$ background. This proposal is attractive in that the boundary term is topological by construction, without invoking additional assumptions such as those employed in \cite{Kurlyand:2022vzv}. However, the overall normalization of the boundary term is fixed only through holographic matching and cannot be determined in a fully intrinsic manner, in contrast to the PST formulation, where the corresponding normalization follows unambiguously from the underlying gauge symmetry structure as emphasized in Subsection \ref{sec:IIB:PST}. We refer the reader to \cite{Chakrabarti:2022jcb} for more details of the construction, and instead briefly comment on a possible connection between their string field theory inspired proposal and our improvement on the PST formulation in Section \ref{sec:discussion}.

\medskip
\noindent\textbf{Clone-Field Formulation}
\medskip

Another line of development achieves a manifestly self-dual and $SL(2, \mathbb{R})$-covariant formulation of Type IIB supergravity, employing a democratic description in which electric and magnetic potentials are treated on an equal footing. Recent realizations of this idea include the clone-field formulation of \cite{Mkrtchyan:2022xrm,Evnin:2023ypu,Hutomo:2025dfx}, while earlier and related approaches to Type II theories, as well as chiral $p$-forms, can be found for example in \cite{DallAgata:1998ahf,Bergshoeff:2001pv,Bandos:2003et,Mkrtchyan:2019opf,Bansal:2021bis,Avetisyan:2021heg,Avetisyan:2022zza,Arvanitakis:2022bnr}. We refer the reader to these works for the detailed construction and below simply quote the bosonic sector of the Euclidean Type IIB action in the clone-field formulation \cite{Mkrtchyan:2022xrm}:  
\begin{align}
	S_\text{IIB,Clone}&=-\fft{1}{2\kappa^2}\int \Big(*R-\fft12d\phi\wedge*d\phi-\fft12e^{-\phi}H_3\wedge*H_3\Big)\nn\\
	&\quad+\fft{1}{4\kappa^2}\int \Big(e^{2\phi}F_1\wedge*F_1+e^{\phi}\tF_3\wedge*\tF_3\Big)\nn\\
	&\quad +\frac{1}{8\kappa^2}\int\Big(\left(F_5+aQ_5\right)\wedge *\left( F_5+aQ_5\right)+2i F_5\wedge aQ_5\nn\\
    &\quad -2i\left(1-i*\right)\left(F_5+aQ_5\right)\wedge \mX_5+\mX_5\wedge*\mX_5\Big) \,.\label{clone-IIB:action}
\end{align}
where $Q_5\equiv dR_4$ and $\mX_5=\tF_5-F_5$ as given in \eqref{IIB:content}. The 4-form $R_4$ is the auxiliary ``clone'' gauge field that implements the democratic doubling of the 5-form sector, while $a$ is the auxiliary scalar field of the PST type, ensuring a covariant implementation of the self-duality constraint. Upon consistently eliminating the auxiliary fields $a$ and $R_4$, the clone-field IIB action \eqref{clone-IIB:action} reduces to the pseudo Euclidean Type IIB action \eqref{IIB:action} modulo the boundary term as required. 

\medskip

To illustrate how the self-duality condition emerges dynamically from the clone-field IIB action \eqref{clone-IIB:action}, we first display the equations of motion for the physical and auxiliary 4-form potentials as
\begin{subequations}
\begin{align}
    \delta_{C_4}:\quad0&=d\left\{*\left(\tF_5+aQ_5\right)+iaQ_5\right\}+iF_3\wedge H_3\,,\label{C}\\
    \delta_{R_4}:\quad0&=d\left[a\left\{*\left(\tF_5+aQ_5\right)-iF_5\right\}\right]+iaF_3\wedge H_3-ida\wedge \mX_5\,.\label{R}
\end{align}\label{clone-IIB:eom}%
\end{subequations}
Combining \eqref{clone-IIB:eom} and making use of the identity \eqref{Identity:3}, one immediately obtains the self-dual relation
\begin{align}
    *\left(\tF_5+aQ_5 \right)&=i\left(\tF_5+aQ_5 \right)\,.\label{new_duality}
\end{align}
We emphasize that the self-duality condition \eqref{new_duality} follows entirely from the equations of motion without requiring a judicious gauge choice, which is in contrast to the PST formulation reviewed in Subsection \ref{sec:IIB:PST}. Another important remark is that, although the auxiliary field strength $Q_5$ can be gauged away \emph{locally} using the gauge symmetry
\begin{align}
    \delta_\ell:&\quad \delta_\ell C_4=\frac{1}{2}a^2\ell_4\,,\quad \delta_\ell R_4=-a\ell_4\,,\quad \delta_\ell (\text{other fields})=0\,,\qquad( d\ell_4=0) \label{CLone-IIB:gg2}
\end{align}
it cannot in general be eliminated \emph{globally} on a generic Type IIB background with non-trivial boundary since the transformation \eqref{CLone-IIB:gg2} constitutes a valid gauge symmetry of the action \eqref{clone-IIB:action} only up to boundary terms. It is therefore natural, in the clone-field formulation, to regard the combination $\tF_5 +aQ_5$ as the genuinely physical self-dual 5-form field strength, rather than $\tF_5$ alone, as advocated in \cite{Mkrtchyan:2022xrm}.

\medskip

Now let us comment on the on-shell value of the clone-field Type IIB action \eqref{clone-IIB:action} evaluated on the $EAdS_5\times S^5$ background presented in Subsection \ref{sec:example:EAdS5}. Upon substituting the background, all contributions cancel except for a single term as
\begin{equation}
    S_\text{IIB,Clone}\Big|_{EAdS_5\times S^5}=\frac{i}{4\kappa^2}\int F_5\wedge aQ_5\,.\label{clone-IIB:action:EAdS5}
\end{equation}
This surviving contribution closely resembles the topological correction to the PST formulation presented in \eqref{PST-IIB:action:2022}. Indeed, in \cite{Mkrtchyan:2022xrm} the expression \eqref{clone-IIB:action:EAdS5} was interpreted as an alternative resolution of the longstanding puzzle that the conventional pseudo-action of Type IIB supergravity yields a vanishing on-shell value on $EAdS_5\times S^5$ \cite{Mkrtchyan:2022xrm}. Compared with the modified PST proposal \eqref{PST-IIB:action:2022}, the clone-field formulation has the advantage of being fully covariant and not tied to a particular background from the outset.

\medskip

There remains, however, an important subtlety before declaring complete success. The clone-field formulation does not, by itself, uniquely fix how the self-dual combination $\tF_5+aQ_5$ should be decomposed into its constituents $\tF_5$ and $Q_5$. For instance, on the $EAdS_5\times S^5$ background with $\tF_5=F_5$, both $F_5$ and $Q_5$ may be taken to be arbitrary closed 5-forms in principle, provided their sum reproduces the required self-dual field strength
\begin{align}
    F_5+aQ_5&=-4(i\epsilon_5+*\epsilon_5)\,,   
\end{align}
where $\epsilon_5$ denotes the unit volume form of $EAdS_5$ throughout the paper. However, the decomposition that yields an on-shell action consistent with holography -- equivalently with the 5d minimal gauged supergravity on-shell action as discussed in Subsection \ref{sec:IIB:PST} -- is in fact naturally singled out as
\begin{align}
    aQ_5 = -4i\epsilon_5\qquad\&\qquad F_5 = -4*\epsilon_5 \, .\label{clone:PST:EAdS5}
\end{align}   
In Subsection \ref{sec:proposal:clone}, we revisit this ambiguity and propose a definitive prescription for the decomposition, guided by the improved PST framework developed in Subsection \ref{sec:proposal:PST}.

\section{Improvement of Type IIB Supergravity}\label{sec:proposal}
In this section, we address the improvement of the Euclidean Type IIB supergravity action in a more general setting. In Subsection \ref{sec:proposal:PST}, we propose a topological term that refines the PST formulation of Euclidean Type IIB supergravity, extending its validity to a broader class of supergravity backgrounds beyond the $M_5\times X_5$ type reviewed in Subsection \ref{sec:IIB:PST}. The consistency of this proposed topological term with holography will be examined in the subsequent Section \ref{sec:example}. In Subsection \ref{sec:proposal:clone}, we revisit the duality symmetric formulation of Euclidean Type IIB supergravity, discussing the intrinsic ambiguity in the decomposition of a self-dual 5-form field strength, which plays a crucial role in holographic comparisons.

\subsection{Refined Topological Term in the PST Formulation}\label{sec:proposal:PST}
As reviewed in Subsection \ref{sec:IIB:PST}, a topological term was added to the original PST-IIB action \eqref{PST-IIB:action}, leading to the improved one \eqref{PST-IIB:action:2022}, by demanding the preservation of the gauge symmetry \eqref{PST-IIB:gg2} in the presence of boundaries. The resulting action was shown to agree exactly on shell with 5d minimal gauged supergravity obtained by dimensional reduction on the internal manifold $S^5$, thereby ensuring consistency with the corresponding holographic observables. This proposal relies crucially on the set of assumptions stated below \eqref{PST-IIB:action:2022}, however, which significantly restrict its domain of applicability and render it inadequate for a broad class of Type IIB AdS backgrounds that arise in the context of holographic duality. 

\medskip

In this subsection, we aim to improve the PST-IIB action \eqref{PST-IIB:action} by introducing an appropriate boundary term that preserves the gauge symmetry \eqref{PST-IIB:gg2}, in close analogy with the modification implemented in \eqref{PST-IIB:action:2022}. Crucially, however, we shall proceed under a minimal set of assumptions, deliberately avoiding those invoked in \eqref{PST-IIB:action:2022}. To this end, we begin by reexamining the gauge transformation \eqref{PST-IIB:gg2} of the PST action \eqref{PST-IIB:action}, under which the variation of the action takes the form
\begin{align}
    \delta_\ell S_{\text{PST}}
    &=\frac{i}{4\kappa^2}\int d\big(\tF_5\wedge \delta_\ell C_4\big)=\frac{i}{4\kappa^2}\int \Big[ d\left(F_5\wedge \delta_\ell C_4\right) + d\left(\mX_5\wedge \delta_\ell C_4\right) \Big]  \label{eq:bdry-id}
\end{align}
with $\mX_5\equiv\tF_5-F_5$ as presented in \eqref{IIB:content}. The variation naturally decomposes into two distinct contributions: the first involves only the 4-form potential $C_4$, whereas the other one depends on the 2-forms through $\mX_5$. Our task is therefore to identify suitable topological terms to be added to the PST action \eqref{PST-IIB:action} whose variation under \eqref{PST-IIB:gg2} precisely cancels the boundary contributions appearing in \eqref{eq:bdry-id}. 

\medskip

To state the conclusion upfront, we propose the following modification of the PST-IIB action \eqref{PST-IIB:action},
\begin{align}
    S_\text{AHJL} = S_\text{PST}  \underbrace{-\fft{i}{4\kappa^2}\int F_{5E}\wedge F_{5NE}}_{\equiv\, S_\text{AHJL}^\text{(top),1}} \underbrace{- \fft{i}{4\kappa^2}\int d \big(\mX_5 \wedge C_4 \big)}_{\equiv\, S_\text{AHJL}^\text{(top),2}}\,, \label{PST-IIB:action:ours}
\end{align}
where we explain below how the two closed forms $F_{5E}$ and $F_{5NE}$ are defined. Note that the proposed correction to the PST action $S_\text{AHJL}^\text{(top)}=S_\text{AHJL}^\text{(top),1}+S_\text{AHJL}^\text{(top),2}$ is purely topological and therefore does not modify the bulk equations of motion as required. In the following, we provide a detailed justification of the proposal \eqref{PST-IIB:action:ours}. 

\medskip

The second topological correction $S_\text{AHJL}^\text{(top),2}$ in \eqref{PST-IIB:action:ours} is engineered to precisely cancel the second term in \eqref{eq:bdry-id} generated by the variation of the original PST action \eqref{PST-IIB:action} under the gauge transformation \eqref{PST-IIB:gg2}.\footnote{The inclusion of this term may equivalently be viewed as a modification of the Chern–Simons sector of the PST-IIB action in \eqref{PST-IIB:action}, namely the last line of \eqref{IIB:action}. From this perspective, if one instead adopts a different convention for the Chern–Simons term from the outset, the variation \eqref{eq:bdry-id} contains only the contribution depending explicitly on the 4-form potential $C_4$. In that case, the only additional topological term required is the purely $C_4$-dependent one, $S_\text{AHJL}^\text{(top),1}$. \label{foot:CS-redef}} At this stage, no further assumptions are required.

\medskip

The first topological correction $S_\text{AHJL}^\text{(top),1}$ in \eqref{PST-IIB:action:ours}, on the other hand, is introduced to cancel the first term in \eqref{eq:bdry-id} under the gauge transformation \eqref{PST-IIB:gg2}. Its definition relies on decomposing the closed 5-form $F_5$ as
\begin{align}
    F_5=F_{5E}+F_{5NE}\,,\label{decompose}
\end{align}
where $F_{5E}$ is a globally well-defined exact form, while $F_{5NE}$ is closed but not globally exact. For this cancellation mechanism to operate, we impose the following three assumptions. Below we have introduced the notation $d_M$ \& $d_X$ to denote the exterior derivatives acting on the external \& internal manifolds respectively, satisfying $d=d_M+d_X$. 
\begin{enumerate}
    \item The 10d background takes a product form $M_d\times X_{10-d}$, possibly equipped with a warped metric, where $M_d$ is a non-compact external manifold with $0\leq d\leq 9$ and $X_{10-d}$ is a compact internal manifold. 

    \item With respect to the decomposition $d=d_M+d_X$, the exact component $F_{5E}$ and the gauge variation of the non-exact component $\delta_\ell F_{5NE}$ obey
    \begin{align}
       d_X F_{5E}=0\qquad \& \qquad \delta_\ell F_{5NE}=d_X\Lambda_4  \label{PST-IIB:gg2:F5}
    \end{align}
    for some globally well-defined 4-form $\Lambda_4$ associated with the gauge parameter $\ell_4$. 

    \item Upon the decomposition \eqref{decompose}, each component of $F_{5E}$ carries a common one-form factor along $M_d$. By contrast, $F_{5NE}$ contains no component involving this particular leg appearing in $F_{5E}$.
\end{enumerate}
Using the above three assumptions, one readily verifies that the first term in the proposal \eqref{PST-IIB:action:ours}, $S^{(\rm top),1}_{\rm AHJL}$, precisely cancels the first term in the gauge variation of the PST-IIB action \eqref{eq:bdry-id} under the gauge transformation \eqref{PST-IIB:gg2} as
\begin{equation}
\begin{split}
    &\delta_\ell S^{(\rm top),1}_{\rm AHJL}+\fft{i}{4\kappa^2}\int d(F_5\wedge \delta_\ell C_4)\\
    &=
    -\fft{i}{4\kappa^2}\int F_{5E}\wedge\delta_\ell F_{5E}
    + 2F_{5E}\wedge\delta_\ell F_{5NE}
    +F_{5NE}\wedge\delta_\ell F_{5NE}\\
    &=
    -\fft{i}{4\kappa^2}\int 
    2F_{5E}\wedge\delta_\ell F_{5NE}
    \qquad(\text{Assumption 3})\\
    &=
    \fft{i}{2\kappa^2}\int 
    \,d_X(F_{5E}\wedge\Lambda_4) \qquad (\text{Assumption 2})\\
    &=0\qquad (\text{Assumption 1})\,.
\end{split}
\end{equation}
Several remarks on the above three assumptions are as follows.
\begin{itemize}
    \item In the first assumption, the case with $d=0$ corresponds to a Type IIB background with a compact 10d manifold, for which the refinement is unnecessary within the PST formulation. It is therefore included merely as a trivial limiting case, and the subsequent assumptions are effectively relevant for $d\geq1$.

    \item In the second assumption, the global exactness of the gauge variation of the 5-form field strength in fact naturally arises from requiring the gauge parameter $\ell_4$ to be globally well-defined, according to \eqref{PST-IIB:gg2}. In other words, we exclude non-trivial large gauge transformations so that the second condition in \eqref{PST-IIB:gg2:F5} holds globally rather than merely patchwise.

    \item The third assumption renders the decomposition \eqref{decompose} unambiguous: the common 1-form factor shared by all components of $F_{5E}$ eliminates the ambiguity of redistributing an arbitrary exact 5-form between $F_{5E}$ and $F_{5NE}$. Furthermore, if the distinguished leg along $M_d$ is chosen to be the Euclidean time differential $d\tau$, one may naturally identify $F_{5E}$ and $F_{5NE}$ with the ``electric'' and ``magnetic'' components of the 5-form respectively, as discussed in Subsection \ref{sec:IIB:PST}. We emphasize that the non-trivial assumptions underlying the decomposition \eqref{decompose} play a crucial role in refining the PST formulation in terms of such electric-magnetic decomposition.

    \item To illustrate the physical relevance of these assumptions, especially the third one, we note that it is naturally satisfied when the external manifold is Euclidean Anti-de Sitter space with trivial de Rham cohomology (except in degree zero). In this case, any closed 5-form component with non-zero degree along $M_d$ is necessarily exact,\footnote{To be more precise, this conclusion holds only if the sub-form within the component of $F_5$, supported along $M_d$, is closed. In various holographic backgrounds where the component of $F_5$ with non-zero degree along $M_d$ is unique, as in the examples of our interest investigated in Section \ref{sec:example}, this condition follows automatically.} implying that the non-exact contribution to the 5-form is exclusively on the internal manifold $X_{10-d}$.\footnote{A consequential observation is that a Type IIB background of the form $EAdS_d\times X_{10-d}$ with $d\geq6$ does not admit a non-trivial non-exact component in the decomposition of the closed 5-form $F_5$ and thereby the first topological correction in \eqref{PST-IIB:action:ours} vanishes identically. We return to this point in Section \ref{sec:discussion}. \label{foot:d>=6}} Consequently, the third assumption holds automatically. This demonstrates that our assumptions are naturally compatible with backgrounds of interest in holographic duality.  
\end{itemize}

\medskip    

In summary, we have established that
\begin{align}
    \delta_\ell S_\text{AHJL}=0\label{IIB-PST:gg:ours}
\end{align}
under the gauge transformation \eqref{PST-IIB:gg2} for a broad class of Type IIB backgrounds satisfying the three assumptions stated above. In Section \ref{sec:example}, we demonstrate that our refined PST-IIB action \eqref{PST-IIB:action:ours} is not only internally consistent within the PST framework -- yielding the self-duality condition for the 5-form field strength for a wider class of Type IIB backgrounds via the gauge symmetry \eqref{IIB-PST:gg:ours} -- but also reproduces, on shell, the corresponding lower-dimensional gauged supergravity results in concrete examples, thereby matching the expected holographic observables.

\medskip

Before turning to explicit AdS/CFT examples, we verify that our proposed topological correction \eqref{PST-IIB:action:ours} reduces to the earlier proposal of \cite{Kurlyand:2022vzv} under the simplifying assumptions stated below \eqref{PST-IIB:action:2022}. To demonstrate this explicitly, we first recall that \cite{Kurlyand:2022vzv} assumes that all other supergravity fields, specifically the 2-form potentials, are absent or pure gauge so that $\mX_5=\tF_5-F_5=0$. Under this assumption, our proposal \eqref{PST-IIB:action:ours} simplifies to
\begin{align}
    S^\text{(top)}_\text{AHJL}\Big|_\text{Assumptions of \cite{Kurlyand:2022vzv}} =   -\fft{i}{4\kappa^2}\int F_{5E}\wedge F_{5NE}\,. 
\end{align}
Next, in the case $d=5$ considered in \cite{Kurlyand:2022vzv}, the third assumption along with $F_5=\tF_5$ unambiguously identifies
\begin{align}
    F_{5E}~~\to~~\tF_{5M}\qquad \&\qquad F_{5NE}~~\to~~\tF_{5X}\,.
\end{align}
Consequently, we obtain
\begin{align}
    S^\text{(top)}_\text{AHJL}\Big|_\text{Assumptions of \cite{Kurlyand:2022vzv}} =   -\fft{i}{4\kappa^2}\int \tF_{5M}\wedge \tF_{5X} = S^\text{(top)}_\text{KT}\,, \label{prev}
\end{align}
thereby establishing the equivalence between \eqref{PST-IIB:action:ours} and \eqref{PST-IIB:action:2022} under the special case of interest. This confirms that the proposal of \cite{Kurlyand:2022vzv} arises as a special case of our more general formulation, valid when the two-form potentials vanish (or are pure gauge) and the geometry factorizes as $M_5 \times X_5$. Our construction, however, extends the applicability of the topological correction to more general Type IIB backgrounds with non-trivial two-form fields and a broader class of product geometries $M_d \times X_{10-d}$. 

\medskip

We close this subsection with a remark on the nature of our proposal. The topological corrections in \eqref{PST-IIB:action:ours} are designed solely to enforce the PST gauge symmetry \eqref{PST-IIB:gg2} in the presence of boundaries, and they do not necessarily constitute the complete boundary structure of the Type IIB action: any further boundary term that is itself invariant under the PST gauge symmetry -- such as a Gibbons-Hawking-York term required for a well-posed variational principle -- is left unfixed by our analysis. From this perspective, it would be very interesting to investigate how the newly proposed boundary terms, especially the Chern-Simons redefinition discussed in Footnote~\ref{foot:CS-redef}, interact with anomaly inflow onto D-branes \cite{Green:1996dd,Cheung:1997az,Minasian:1997mm} or with the holographic renormalization \cite{Henningson:1998gx,deHaro:2000vlm,Bianchi:2001kw,Skenderis:2002wp} that remains subtle for general Type IIB backgrounds \cite{Taylor:2001fe,Taylor:2001pp,Skenderis:2006uy,Skenderis:2007yb,IzquierdoGarcia:2025jyb,Anempodistov:2026dhi}; such studies may provide an independent viewpoint on the proposed boundary terms, and we leave them to future work.

\subsection{Comments on the Boundary Term in the Clone-Field Formalism}\label{sec:proposal:clone}
In this subsection, we comment on the clone-field formulation of Type IIB supergravity reviewed in Subsection \ref{sec:IIB:other}, focusing on a subtle ambiguity that arises in the decomposition of the self-dual 5-form field strength. In that formalism, the self-dual combination $\tF_5+aQ_5$ is introduced at the level of the action, but its splitting into individual pieces, $\tF_5=F_5+\mX_5$ and $aQ_5$, is not uniquely fixed. This freedom translates into an ambiguity in identifying the ``electric'' and ``magnetic'' components of the 5-form field strength and, consequently, in the evaluation of the on-shell action in backgrounds with non-trivial boundary contributions.

\medskip

To clarify this issue, let us compare the bosonic action in the clone-field formulation \eqref{clone-IIB:action} with our refined PST-IIB action \eqref{PST-IIB:action:ours}. A direct inspection shows that the two actions coincide provided we identify the 5-form field strengths according to
\begin{align}
    F_5 \longleftrightarrow F_{5NE}
    \qquad\&\qquad aQ \longleftrightarrow F_{5E} \,, \label{clone:PST}
\end{align}
where it is crucial to distinguish the $F_5$ appearing in the clone-field formulation from the total 5-form $F_5=F_{5E}+F_{5NE}$ in the refined PST description. In other words, once the ambiguity in the clone-field decomposition is fixed by the prescription \eqref{clone:PST}, the clone-field and refined PST formalisms become equivalent, at least as far as the physics determined by the bosonic action is concerned.

A particularly important observable in this context is the on-shell action, which in holographic applications encodes the leading large-$N$ behavior of the dual field theory observable. Under the identification \eqref{clone:PST}, the clone-field and refined PST formalisms yield precisely the same non-vanishing on-shell action for the class of Type IIB backgrounds considered in this work. In the special case of the $EAdS_5\times S^5$ background where \eqref{clone:PST} reduces to the specific prescription \eqref{clone:PST:EAdS5}, the resulting on-shell action indeed reproduces the expected holographic quantity, as briefly discussed already in Subsection \ref{sec:IIB:other}. In Section \ref{sec:example}, we will extend this analysis to more non-trivial backgrounds and demonstrate that the resulting on-shell action agrees with the expectation from the lower-dimensional gauged supergravity, thereby providing further confirmation of its consistency with holography. From this perspective, \eqref{clone:PST} emerges as the most natural bridge between the clone-field and refined PST formulations for the broad class of Type IIB backgrounds specified in Subsection \ref{sec:proposal:PST}.

\medskip

We emphasize, however, that this fixing of the ambiguity has been motivated indirectly through comparison with the refined PST formulation rather than derived intrinsically within the clone-field framework itself. Providing an internal justification of the prescription \eqref{clone:PST} purely from the structure of the clone-field formalism remains an open problem. We leave a systematic investigation of this issue to future work.

\section{Type IIB Supergravity On-shell Action}\label{sec:example}
In this section, we evaluate the improved PST action for 10d Type IIB supergravity \eqref{PST-IIB:action:ours} on shell for three distinct holographic backgrounds -- $EAdS_5\times S^5$, the Lunin-Maldacena (LM) background, and the $AdS_4$ $S$-fold solution -- and compare the results with those obtained in the corresponding lower-dimensional gauged supergravity descriptions, which have been shown to reproduce the expected holographic observables. The latter two examples play a particularly important role, as they provide non-trivial tests of our proposal for the refined PST-IIB action \eqref{PST-IIB:action:ours}, extending the earlier formulation \eqref{PST-IIB:action:2022} presented in \cite{Kurlyand:2022vzv}, which was primarily examined in the $EAdS_5\times S^5$ vacuum background. In particular,
\begin{itemize}
    \item although the LM background is topologically equivalent to $EAdS_5\times S^5$, it involves non-trivial 2-form potentials,

    \item the $S$-fold solution features external and internal manifolds of different dimensions,
\end{itemize}
where both aspects fall outside the restrictive assumptions underlying the previous formulation and are naturally accommodated in our refined PST-IIB action \eqref{PST-IIB:action:ours}.

\medskip

Before turning to the analysis of specific backgrounds, we first substitute the PST-IIB equations of motion \eqref{PST-IIB:eom} into the PST action \eqref{PST-IIB:action} and simplify the resulting general on-shell expression. To be specific, using the Ricci scalar determined from the Einstein equations \eqref{PST-IIB:eom:g}
\begin{align}
    R&=\fft12\partial^\mu\phi\partial_\mu\phi+\fft12e^{2\phi}\partial^\mu C_0\partial_\mu C_0+\frac{1}{24}\left(e^{-\phi}H^{\mu\rho\sigma}H_{\mu\rho\sigma}+e^{\phi}\tF^{\mu\rho\sigma}\tF_{\mu\rho\sigma}\right)\nn\\
    &=\frac{1}{2}|{d\phi}|^2+\frac{1}{2}e^{2\phi}|F_1|^2+\frac{1}{4}\left(e^{-\phi}|H_3|^2+e^{\phi}|\tF_3|^2\right)\label{Ricci}
\end{align}
together with the self-duality constraint for the 5-form field strength $\tF_5$, the PST action \eqref{PST-IIB:action} reduces on shell to
\begin{align}
    S^{\text{On-Shell}}_\text{PST}   
    &=-\frac{1}{2\kappa^2	}\int \bigg(-\fft14e^{-\phi}H_3\wedge*H_3-\fft{1}{4}e^{\phi}\tF_3\wedge*\tF_3-\fft{i}{2} C_4\wedge H_3\wedge F_3\bigg) \label{tot_derivative}\\
    &=\fft{1}{8\kappa^2}\int d\left( B_2\wedge e^{-\phi} * H_3 +e^{\phi}C_2\wedge *\tF_3 -e^\phi C_0 B_2\wedge *\tF_3\right)+\fft{i}{4\kappa^2}\int d\left(C_4 \wedge \mX_5 \right)\,.\nn
\end{align}
Then, incorporating the additional topological correction introduced in the proposal \eqref{PST-IIB:action:ours}, we obtain the on-shell expression for the refined PST-IIB action:
\begin{align}
    S^{\text{On-Shell}}_{\text{AHJL}}&\equiv S^{\text{On-Shell}}_\text{PST}+ S^{\text{(top),1}}_{\text{AHJL}}+S^{\text{(top),2}}_{\text{AHJL}}\label{gen_on_shell}\\
    &=\fft{1}{8\kappa^2}\int d\left( B_2\wedge e^{-\phi} * H_3 +e^{\phi}C_2\wedge *\tF_3 -e^\phi C_0 B_2\wedge *\tF_3\right)-\frac{i}{4\kappa^2}\int F_{5E}\wedge F_{5NE}\,. \nn
\end{align}
Importantly, the inclusion of the topological terms does not alter the bulk equations of motion, which justifies the general on-shell expression \eqref{gen_on_shell}. In the remainder of this section, we evaluate the refined PST-IIB on-shell action \eqref{gen_on_shell} for the holographic Type IIB backgrounds mentioned above, where the radius of the EAdS geometry is set to unity for notational convenience.

\subsection{\texorpdfstring{$EAdS_5\times S^5$}{EAdS5 x S5} Solution}\label{sec:example:EAdS5}
In this subsection, we revisit the well-known $EAdS_5\times S^5$ background, which was already analyzed in \cite{Kurlyand:2022vzv}. As discussed in Subsection \ref{sec:proposal:PST}, since our refined PST-IIB action \eqref{PST-IIB:action:ours} reduces to their formulation \eqref{PST-IIB:action:2022} on the $EAdS_5\times S^5$ vacuum, the final conclusion remains unchanged: the on-shell refined PST-IIB action \eqref{gen_on_shell} yields a non-vanishing contribution arising entirely from the topological correction, which precisely reproduces the on-shell action of the 5d gauged supergravity. Our purpose here is therefore not to obtain a new result, but rather to illustrate how our terminology and decomposition apply in the simplest example and to clarify the mechanism underlying our proposal.

\medskip

The $EAdS_5 \times S^5$ vacuum solution to the PST-IIB equations of motion \eqref{PST-IIB:eom} is given by
\begin{subequations}
\begin{align}
    ds^2_{10}&=ds^2_{EAdS_5}+ds^2_{S^5}\label{EAdS5:metric}\,,\\
    \tF_5=F_5&=-4(i\epsilon_5+*\epsilon_5)\label{EAdS5:F5}\,,\\
    C_0,\phi&=\text{const}\,,\\
    B_2=C_2&=0\,,
\end{align}\label{EAdS5}%
\end{subequations}
where $\epsilon_5$ denotes the unit volume form of $EAdS_5$. Substituting the vacuum configuration \eqref{EAdS5} into the refined PST-IIB on-shell action \eqref{gen_on_shell}, we find that the first three terms in the integrand identically vanish, leaving only the topological contribution as
\begin{align}
   S^{\text{On-Shell}}_{\text{AHJL}}\Big|_{\eqref{EAdS5}}=-\frac{i}{4\kappa^2}\int F_{5E}\wedge F_{5NE}\,.
\end{align}
Moreover, the third assumption introduced in the general formulation of Subsection \ref{sec:proposal:PST} uniquely fixes the decomposition of the 5-form field strength \eqref{EAdS5:F5} into its exact and non-exact components as
\begin{align}
    F_{5E}=-4i\epsilon_5 \qquad\&\qquad F_{5NE}=-4*\epsilon_5\,.
\end{align}
Consequently, the refined PST-IIB on-shell action \eqref{gen_on_shell} evaluated on the $EAdS_5 \times S^5$ background \eqref{EAdS5} becomes
\begin{align}
    S^{\text{On-Shell}}_{\text{AHJL}}\Big|_{\eqref{EAdS5}}&=\fft{4}{\kappa^2}\text{Vol}_{EAdS_5}\text{Vol}_{S^5}\,.\label{S10:EAdS5}
\end{align}

\medskip

On the other hand, the 5d gauged supergravity on-shell action for the $EAdS_5$ background is given by
\begin{align}
    S_5\Big|_{EAdS_5}=-\fft{1}{16\pi G_N^{(5)}}\int d^5x_{EAdS_5}\,\sqrt{g_{EAdS_5}}\Big[R^{(5)}+12+\cdots\Big]=\fft{\Vol_{EAdS_5}}{2\pi G_N^{(5)}}\,,\label{S5:EAdS5:G5}
\end{align}
where the ellipsis denotes contributions from additional fields in the 5d gauged supergravity arising from consistent truncation of Type IIB supergravity on $S^5$ \cite{Cvetic:1999xp,Cvetic:2000nc,Baguet:2015sma}, all of which vanish on shell for the present purely gravitational background. To proceed, we employ the relation between the 5d Newton constant and the 10d gravitational coupling obtained via dimensional reduction under the consistent truncation on $S^5$:
\begin{align}
    -\fft{1}{2\kappa^2} \int d^{10}x \, \sqrt{G_{10}} \, R^{(10)} &= -\fft{1}{16\pi G_N^{(5)}} \int d^5x_{EAdS_5}\, \sqrt{g_{EAdS_5}}\,\Big[R^{(5)}+\cdots\Big]\nn\\
    \to\qquad \fft{\Vol_{S^5}}{2\kappa^2}&=\fft{1}{16\pi G_N^{(5)}}\,.\label{G5:G10}
\end{align}
Substituting this relation into the 5d on-shell action \eqref{S5:EAdS5:G5}, we obtain
\begin{align}
    S_5\Big|_{EAdS_5}=\fft{4}{\kappa^2}\text{Vol}_{EAdS_5}\text{Vol}_{S^5}\,,\label{S5:EAdS5}
\end{align}
which precisely matches the refined PST-IIB on-shell action \eqref{S10:EAdS5} computed directly in the 10d parent theory.

\medskip

To compare the on-shell actions \eqref{S10:EAdS5} and \eqref{S5:EAdS5} with the $S^4$ free energy of the dual $\mN=4$ $\mathrm{SU}(N)$ SYM theory, one must in principle perform holographic renormalization to regulate the divergent volume of the $EAdS_5$ geometry \cite{Henningson:1998gx,deHaro:2000vlm,Skenderis:2002wp}. However, this matching has already been established: holographic renormalization of the 5d action reproduces the field theory result \cite{Henningson:1998gx}. Since the 10d on-shell action derived above is identical to the 5d result -- most notably sharing the same divergent factor proportional to $\Vol_{EAdS_5}$ -- the renormalized value of the 10d action necessarily coincides with the dual field theory observable. For this reason, we do not repeat the holographic renormalization procedure here, and the same reasoning will be applied in the subsequent subsections for other Type IIB backgrounds.

\subsection{Lunin-Maldacena Solution}\label{sec:example:LM}
In this subsection, we analyze the on-shell value of the refined PST-IIB supergravity action \eqref{gen_on_shell} for the LM background \cite{Lunin:2005jy}, which describes the $\mN=1$ exactly marginal $\beta$-deformation of the $\mN=4$ SYM theory \cite{Leigh:1995ep,Aharony:2002hx} through holographic duality. 

\medskip

The LM solution to the PST-IIB equations of motion \eqref{PST-IIB:eom} takes the form \cite{Lunin:2005jy}
\begin{subequations}
\begin{align}
    ds^2&= G^{- 1/4} \left[ ds^2_{EAdS_5} + \sum_i( d\mu_i^2 + G \mu_i^2 d\phi_i^2)  + 9(\gamma^2+\sigma^2) G\mu_1^2 \mu_2^2 \mu_3^2 d \psi^2 \right]\,, \label{LM:metric}\\
    e^{-\phi} &= G^{-1/2} H^{-1}\,, \qquad C_0 =\gamma \sigma g_{0,E} H^{-1}\,, \\
    B_2 &= \ \gamma G w_2 - 12 \sigma w_1 \wedge d \psi \,, \qquad
    C_2 = \ -\sigma G w_2 - 12 \gamma w_1 \wedge d \psi \,, \\
    \tF_5 &=-4(i+*)\epsilon_5\,,\label{LM:F5}
\end{align}\label{LM}%
\end{subequations}
where $\epsilon_5$ again denotes the volume form of unit $EAdS_5$ and we have introduced 
\begin{align}
    G^{-1} & =  1 + (\gamma^2 + \sigma^2) g_{0,E}\,, \qquad H = 1 + \sigma^2 g_{0,E}\,,\qquad g_{0,E} =\mu_1^2 \mu_2^2 + \mu_2^2 \mu_3^2 + \mu_3^2 \mu_1^2\,,\nn\\
    dw_1 &=\mu_1\mu_2\mu_3*_21
    \,,\nn\\
    w_2& = \mu_1^2 \mu_2^2 d\phi_1\wedge d\phi_2 + \mu_2^2 \mu_3^2 d\phi_2\wedge d\phi_3 + \mu_3^2 \mu_1^2 d\phi_3\wedge d\phi_1\,.\label{LM:GHg}
\end{align}
In this construction, the five-sphere is expressed as a $T^3$ fibration over $S^2$, where the coordinates $\{\phi_i\}$ parameterize the torus and $\{\mu_i\}$ denote directional cosines satisfying the constraint $\sum_i \mu_i^2 = 1$. The combination $\psi\equiv(\phi_1+\phi_2+\phi_3)/3$ corresponds to the diagonal isometry direction associated with the $\mathrm{U}(1)_R$ symmetry. The resulting solution depends on two real deformation parameters, $\gamma$ and $\sigma$.

\medskip

Now we substitute the above LM solution \eqref{LM} back into the on-shell action given in \eqref{gen_on_shell}. We first observe that the first total-derivative term in \eqref{gen_on_shell} vanishes identically for the LM background: although the 9-forms appearing inside the exterior derivative are non-zero due to the presence of nontrivial two-form potentials, each of them carries a component along the radial direction and therefore the corresponding boundary integral obtained via Stokes' theorem evaluates to zero at the radial cutoff. The only remaining contribution therefore arises from the final term in \eqref{gen_on_shell}, which corresponds to a part of the topological correction,
\begin{align}
    S^{\text{On-Shell}}_{\text{AHJL}}\Big|_{\eqref{LM}}=-\fft{i}{4\kappa^2}\int F_{5E}\wedge F_{5NE}\,.
\end{align}
Importantly, this closed 5-form $F_5$ should be distinguished from the self-dual field strength $\tF_5$ and can be read off from the background \eqref{LM:F5} as
\begin{align}
    F_5 =-4(i+G^{-1}*)\epsilon_5\,,
\end{align}
where the exact and non-exact components are determined according to the third assumption stated in Subsection \ref{sec:proposal:PST} as
\begin{align}
    F_{5E}=-4i\epsilon_5 \qquad \& \qquad F_{5NE}=-4G^{-1}*\epsilon_5\,.\label{LM:F5:decomposition}
\end{align}
This example illustrates that the 5-form decomposition rule in Subsection~\ref{sec:proposal:PST} works unambiguously for a geometry involving a non-trivial warp factor, since the rule is fixed by the leg structure of the 5-form rather than by the warp factor itself.\footnote{Interestingly, the warp factor happens to appear only in the non-exact component, so $F_{5E}$ and $F_{5NE}$ are supported entirely on the external and internal manifolds, respectively. However, establishing a solution-independent principle for the distribution of the warp factor between $F_{5E}$ and $F_{5NE}$ in generic warped (and possibly fibred) Type IIB backgrounds would require a systematic analysis of the equations of motion, which we leave as an interesting future direction.} The on-shell value of the refined PST-IIB action \eqref{gen_on_shell} for the LM background \eqref{LM} then follows as 
\begin{align}
    S^{\text{On-Shell}}_{\text{AHJL}}\Big|_{\eqref{LM}}=-\frac{i}{4\kappa^2}\int\left(-4i\epsilon_5\right)\wedge\left(-4G^{-1}*\epsilon_5\right)=\frac{4}{\kappa^2}\Vol_{EAdS_5}\Vol_{S^5}\,.\label{S10:LM}
\end{align}
Notably, the final expression is independent of the deformation parameters $\gamma$ and $\sigma$. This cancellation follows from the additional factor of $G$ appearing in the non-exact component given in \eqref{LM:F5:decomposition}, which arises in the decomposition of the closed 5-form $F_5$, but not in the self-dual field strength $\tF_5$. As will be seen below, this feature plays a crucial role in achieving agreement with the lower-dimensional supergravity description. 

\medskip

Next we consider the on-shell action of 5d minimal gauged supergravity obtained from the consistent truncation of Type IIB supergravity on the Lunin-Maldacena background \cite{Liu:2019cea}. The resulting expression coincides exactly with that of the undeformed $EAdS_5$ vacuum, namely
\begin{align}
    S_5\Big|_{EAdS_5}&=-\fft{1}{16\pi G_N^{(5)}}\int d^5x_{EAdS_5}\,\sqrt{g_{EAdS_5}}\Big[R^{(5)}+12+\cdots\Big]=\fft{\Vol_{EAdS_5}}{2\pi G_N^{(5)}}\nn\\
    &=\fft{4}{\kappa^2}\text{Vol}_{EAdS_5}\text{Vol}_{S^5}\,,\label{S5:LM}
\end{align}
where the second line follows from the relation between the 5d and 10d gravitational couplings \eqref{G5:G10}, which remains valid for the $\beta$-deformed LM geometry \eqref{LM:metric}. Importantly, the volume of the genuinely deformed internal manifold -- the squashed five-sphere -- does not appear in this relation due to the specific manner in which the warp factor enters the metric as demonstrated explicitly in \cite{Freedman:2005cg}. As a consequence, the lower-dimensional supergravity on-shell action agrees precisely with the refined PST-IIB on-shell action \eqref{S10:LM}.

\medskip

Concerning holography, the standard holographic comparison discussed in Subsection \ref{sec:example:EAdS5} between the on-shell action of $EAdS_5$ and the large $N$ free energy of the dual CFT on $S^4$ extends directly to the Lunin-Maldacena background without modification, since the $\beta$-deformation is exactly marginal \cite{Lunin:2005jy,Freedman:2005cg}. Beyond this, one may further enrich this correspondence by considering $EAdS_5$ with alternative supersymmetric conformal boundaries. In particular, when the conformal boundary is taken to be $S^1\times S^3$, the gravitational on-shell action is expected to reproduce the supersymmetric Casimir energy appearing in the supersymmetric partition function of the dual theory \cite{Assel:2015nca,Bobev:2015kza,Martelli:2015kuk,Lorenzen:2014pna,Brunner:2016nyk}. The holographic realization of this relation was established in 5d gauged supergravity, where the correct matching requires supplementing the standard holographic renormalization procedure with supersymmetry-preserving boundary terms \cite{BenettiGenolini:2016qwm,BenettiGenolini:2016tsn}. The analysis presented in this subsection shows that the same holographic comparison can in fact be implemented directly between the 10d parent theory on-shell action and the supersymmetric Casimir energy of dual field theories.

\subsection{\texorpdfstring{$EAdS_4 \times S^1 \times S^5$}{EAdS4 x S1 x S5} $S$-fold Solution}\label{sec:example:S-fold}
We now turn to the $S$-fold solution constructed in \cite{Inverso:2016eet} -- see also \cite{Giambrone:2021wsm,Guarino:2022tlw} for subsequent analyses and reviews -- and analyze the corresponding on-shell value of the refined PST-IIB supergravity action \eqref{gen_on_shell}. 

\medskip

In Euclidean signature, the $S$-fold background of interest takes the form \cite{Giambrone:2021wsm} 
\begin{subequations}
\begin{align}
    ds_{10}^2 &= \Delta^{-1}\left( \frac{1}{2}ds^2_{EAdS_4} + d\eta^2 + d\alpha^2 + \frac{\cos^2\alpha}{2 + \cos(2\alpha)}\, d\Omega_1^2 + \frac{\sin^2\alpha}{2 - \cos(2\alpha)}\, d\Omega_2^2 \right),\label{S-fold:metric}\\
    e^{\Phi} &= \sqrt{2}\, e^{-2\eta} \frac{2 - \cos(2\alpha)}{7 - \cos(4\alpha)}\,, \qquad C_0 =0\,, \label{S-fold:dilaton}\\
    B_2 &= -2\sqrt{2}\, e^{-\eta} \frac{\cos^3\alpha}{2 + \cos(2\alpha)}\, \text{vol}_{\Omega_1} \,, \qquad
    C_2 = -2\sqrt{2}\, e^{\eta} \frac{\sin^3\alpha}{2 - \cos(2\alpha)}\, \text{vol}_{\Omega_2}\,, \\
    C_4 &= \frac{3}{2}\, \omega_3 \wedge \left(d\eta + \frac{2}{3}\, \sin(2\alpha)\, d\alpha\right)- \frac{1}{2}\, f(\alpha)\, d\alpha \wedge (A_1 \wedge \text{vol}_{\Omega_2} + \text{vol}_{\Omega_1} \wedge A_2),\label{S-fold:C4}
\end{align}\label{S-fold}%
\end{subequations}
where we have defined
\begin{align}
    \Delta^{-4} &= 4 - \cos^2(2\alpha)\,, \qquad d\omega_3 = -i\text{vol}_{EAdS_4}\,,\qquad A_i = -\cos\theta_i\, d\varphi'_i\,,\nn\\
    d\Omega_i^2 &= d\theta_i^2 + \sin^2\theta_i\, d{\varphi'}_i^2 \quad \text{with} \quad d{\varphi}'_i = d\varphi_i + \chi_i\, d\eta\,,\nn\\
    f(\alpha) &= \sin^2(2\alpha)\frac{\cos(4\alpha)-55}{(7 - \cos(4\alpha))^2}\,.\label{S-fold:falpha}
\end{align}
As is manifest from the above solution, the internal space is topologically equivalent to $S^1_{\eta} \times S^5$. The five-sphere $S^5$ is described as a fibration
\begin{equation}
    S^5 = I \times S^2_1 \times S^2_2 \,,
\label{eq:S5decomp}
\end{equation}
where $I$ denotes an interval parameterized by the angular coordinate $\alpha\in[0,\fft\pi2]$. 

\medskip

The refined PST-IIB on-shell action \eqref{gen_on_shell} evaluated on the $S$-fold solution \eqref{S-fold} is simply given by
\begin{align}
    S^{\text{On-Shell}}_{\text{AHJL}}\Big|_{\eqref{S-fold}}=-\frac{i}{4\kappa^2}\int \bigg(F_{5E}\wedge F_{5NE}\bigg)\,,\label{S10:S-fold:1}
\end{align}  
since the first three total-derivative terms in \eqref{gen_on_shell} vanish as in the previous examples. The closed 5-form $F_5$ can be read off from \eqref{S-fold:C4}, and we decompose it into exact and non-exact components following the third assumption of the general formulation presented in Subsection \ref{sec:proposal:PST} as
\begin{align}
    F_5&=F_{5E}+F_{5NE}\,, \\
    F_{5E}&=-\frac{3i}{2} \vol_{EAdS_4}\wedge \Big(d\eta+\frac{2}{3}\sin{2\alpha}d\alpha\Big)\,,\nn\\
    F_{5NE}&= f(\alpha)\, d\alpha \wedge \vol_{\Omega_1}\wedge \vol_{\Omega_2}\,.\nn
\end{align}
Substituting this decomposition into \eqref{S10:S-fold:1}, the refined PST-IIB on-shell action can be evaluated explicitly as
\begin{align}
    S^{\text{On-Shell}}_{\text{AHJL}}\Big|_{\eqref{S-fold}}&=-\frac{i}{4\kappa^2}\int \Big(-\frac{3i}{2} \vol_{EAdS_4}\Big)\wedge \Big(d\eta+\frac{2}{3}\sin{2\alpha}d\alpha\Big)\wedge \, f(\alpha)\, d\alpha \wedge \vol_{\Omega_1}\wedge \vol_{\Omega_2} \nn\\
    &=\frac{3\pi^2}{16\kappa^2}\Vol_{S_1^2}\Vol_{S_2^2}\Vol_{EAdS_4} \,.\label{S10:S-fold:2}
\end{align}

\medskip

The $S$-fold solution described above admits a consistent truncation of Type IIB supergravity on the internal space $S^1 \times S^5$ to 4d maximal $\mN=8$ dyonically gauged supergravity \cite{Inverso:2016eet}, with a further truncation to a specific subsector of this maximal theory identified in \cite{Guarino:2024gke}. In this context, we analyze the on-shell action of 4d gauged supergravity evaluated on the $EAdS_4$ vacuum \cite{Gallerati:2014xra}, which takes the form
\begin{align}
    S_4\Big|_{EAdS_4}&=-\fft{1}{16\pi G_N^{(4)}}\int d^4x_{EAdS_4}\,\sqrt{g_{EAdS_4}}\Big[R^{(4)}+6+\cdots\Big]=\fft{3\Vol_{EAdS_4}}{8\pi G_N^{(4)}}\,,\label{S4:S-fold:G4}
\end{align}
where the ellipsis represents contributions involving other 4d supergravity fields vanishing on-shell at the vacuum. The result \eqref{S4:S-fold:G4} can be rewritten in terms of the 10d gravitational coupling by employing the following relation obtained from dimensional reduction,
\begin{align}
    -\fft{1}{2\kappa^2} \int d^{10}x \, \sqrt{G_{10}} \, R^{(10)} &= -\fft{1}{16\pi G_N^{(4)}} \int d^4x_{EAdS_4}\, \sqrt{g_{EAdS_4}}\,\Big[R^{(4)}+\cdots\Big]\nn\\
    \to\qquad \fft{\pi^2\Vol_{S_1^2}\Vol_{S_2^2}}{32\kappa^2}&=\fft{1}{16\pi G_N^{(4)}}\,,\label{G4:G10}
\end{align}
which follows from substituting the $S$-fold metric \eqref{S-fold:metric} into the left hand side. The resulting expression for the on-shell action becomes
\begin{align}
    S_4\Big|_{EAdS_4}&=\frac{3\pi^2}{16\kappa^2}\Vol_{S_1^2}\Vol_{S_2^2}\Vol_{EAdS_4}\,,\label{S4:S-fold}
\end{align}
which precisely agrees with the direct computation \eqref{S10:S-fold:2} based on the refined PST-IIB on-shell action \eqref{gen_on_shell}.

\medskip

On the field theory side, the dual 3d SCFT was constructed \cite{Assel:2018vtq} via a suitable $S$-folding of a class of supersymmetric Janus supergravity solutions \cite{DHoker:2007zhm,DHoker:2007hhe} dual to Janus interface CFTs \cite{DHoker:2006qeo,Gaiotto:2008sd}. In the same work, a nontrivial test of the holographic correspondence was carried out by matching the on-shell action of the 4d gauged supergravity previously analyzed in \cite{Assel:2012cp,Assel:2012cj} with the $S^3$ free energy computed using supersymmetric localization. The refined PST-IIB formulation, which reproduces the lower-dimensional supergravity result at the level of the on-shell action for the $S$-fold background as described above, now enables this comparison to be performed directly in the 10d parent theory as in the previous examples. More broadly, the refined PST-IIB description developed here should provide a useful starting point to revisit quantitative tests of holography for a wide range of Janus-, $J$-fold, $S$-fold related backgrounds directly within the 10d parent theory, involving Sasaki-Einstein $J$-fold constructions, holographic interfaces in $\mN=4$ SYM, the conformal manifolds of 3d $S$-fold SCFTs, and their associated marginal deformations and RG flows \cite{Bobev:2019jbi,Guarino:2019oct,Bobev:2020fon,Bobev:2021yya,Arav:2021tpk,Bobev:2023bxs}.

\section{Discussions}\label{sec:discussion}
We have reviewed the PST formulation of the bosonic sector of Type IIB supergravity, in which the self-duality of the 5-form field strength follows directly from a manifestly covariant action, including a consistent treatment in Euclidean signature, and briefly outlined two alternative approaches with the same objective -- a string-field-theory-motivated construction and the clone-field formalism. We explained the necessity of refining the PST framework for Type IIB backgrounds with boundaries, as commonly encountered in holographic settings. While an initial refinement was proposed in \cite{Kurlyand:2022vzv}, it relied on rather restrictive assumptions about the background geometry and field content. We have introduced a less restrictive refinement applicable to a broader class of Type IIB backgrounds, including various EAdS geometries central to holography, and demonstrated in two non-trivial examples -- the Lunin-Maldacena solution and $S$-fold geometries -- that our proposal yields a non-vanishing on-shell action consistent with holographic expectations. We also briefly discussed how the action derived from the clone-field formalism can be made compatible with holography by resolving the ambiguity in the splitting of the self-dual 5-form field strength. 

\medskip

Below we outline several open problems that naturally arise from our analysis. 

\medskip

A fully general refinement applicable to arbitrary Type IIB solutions remains to be developed. As a first step, it would be instructive to analyze explicit examples of Type IIB backgrounds containing an $EAdS_d$ geometry with $d\geq6$. For such backgrounds, $S_\text{AHJL}^\text{(top),1}$ vanishes structurally and $S_\text{AHJL}^\text{(top),2}$ amounts to a Chern-Simons sector redefinition as noted in Footnotes~\ref{foot:d>=6} and \ref{foot:CS-redef} respectively; the proposal \eqref{PST-IIB:action:ours} therefore becomes equivalent to the original PST action \eqref{PST-IIB:action} with a redefined Chern-Simons sector, while PST gauge invariance is still ensured. Explicit examples are then needed to determine whether the resulting on-shell action correctly reproduces the corresponding lower-dimensional gauged supergravity result, or whether a further structural refinement of \eqref{PST-IIB:action:ours} is required. Prominent examples include the warped $AdS_6\times S^2$ solutions of \cite{DHoker:2016ujz,DHoker:2017mds,DHoker:2017zwj}, which admit consistent truncations to 6d F(4) gauged supergravity \cite{Hong:2018amk,Malek:2018zcz}. Secondly, it is desirable to relax the structural assumptions imposed on the Type IIB background geometry in this work -- most notably the constraints on the factorization between the external and internal manifolds. Indeed, many holographic Type IIB backgrounds of interest involve non-trivial fibrations between these sectors, as occurs for instance in AdS black hole geometries. Extending the refined PST-IIB formalism to accommodate such configurations therefore constitutes a natural and worthwhile direction for future investigation. More broadly, establishing a refined PST-IIB action that uniformly reproduces the correct on-shell action for all known holographic Type IIB backgrounds, possibly through further improvements to the proposal \eqref{PST-IIB:action:ours}, would furnish a genuine first-principles justification of the formulation.

Extension to other 10,11d supergravity theories would be another intriguing direction to explore. Unlike Type IIB, the other 10,11d supergravities do not contain manifestly self-dual fields, so the precise structure of the refinement may differ. Nevertheless, duality-symmetric formulations exist in eleven dimensions, notably the PST-type construction of \cite{Bandos:1997gd} (see also \cite{Sorokin:1998kf,Bandos:2003et}) and more recent developments such as \cite{Evnin:2023ypu,Rosabal:2025zry}. It would be an interesting non-trivial task to determine whether these formulations reproduce the correct on-shell action compatible with lower-dimensional gauged supergravity truncations and holographic expectations. A pragmatic prescription -- reversing the sign of the flux contribution for ``electric'' solutions so as to obtain an on-shell action consistent with holography -- has already been advocated in the context of eleven-dimensional supergravity \cite{Beccaria:2023hhi,Beccaria:2023ujc}. However, a systematic derivation of this rule from a fundamental formulation of supergravity, such as a PST or clone-field framework, is still lacking. In particular, it would be conceptually preferable to derive the necessary boundary or topological contributions intrinsically from the supergravity side, rather than inferring them from holographic matching. 

The precise relation between the refined PST/clone-field formulations and string-field-theory-motivated approach calls for further clarification. A structural comparison between the three formulations was first initiated in \cite{Evnin:2022kqn}, and extending that analysis to incorporate the refined proposal presented in Section \ref{sec:proposal} could help clarify whether these approaches constitute genuinely distinct formulations or instead represent different realizations of a common underlying structure. In this regard, a recent work \cite{Hutomo:2025dfx} has clarified the equivalence between the PST and clone-field formulations modulo boundary terms for non-linear chiral $p$-form theories in $D=2p+2$, while also establishing a connection to the earlier duality-covariant construction of 4d self-dual theories \cite{Ivanov:2014nya}. On the other hand, the off-shell connection between the refined PST/clone-field formulations and the string-field-theory-motivated framework, which introduces boundary terms involving decoupled unphysical fields as briefly discussed in Subsection \ref{sec:IIB:other}, remains far from transparent. At the level of the on-shell action, however, one may anticipate a more concrete correspondence: indeed, the relations between the supergravity fields and their string-field-theory counterparts have been consistently established using the string-field-theory equations of motion \cite{Mamade:2025jbs}, opening a window to investigate how the topological correction proposed in \eqref{PST-IIB:action:ours} relates to the corresponding boundary term on the string-field-theory side. See also \cite{Arvanitakis:2022bnr,Lambert:2023qgs,Hull:2023dgp,Evnin:2023ypu,Hull:2025mtb,Hull:2026osk} for related recent discussions, although this list is by no means exhaustive.

Lastly, exploring and justifying the proposed topological correction beyond the on-shell action level is essential for a deeper understanding of the Type IIB supergravity path integral and its comparison with dual field theory observables beyond the semi-classical approximation. In the present work, we have shown that the topological correction modifies the on-shell value of the supergravity action, thereby altering the leading saddle point contribution to the Type IIB supergravity path integral, which turns out to be consistent with the corresponding large $N$ quantity in the dual field theory. This provides non-trivial support for the refinement at the classical level. A natural next step is to determine whether this correction also influences the one-loop structure of the supergravity path integral. If the topological term contributes non-trivially at this level, one should analyze its impact on the subleading $1/N$ corrections of the dual theory and test the refined formulation within the framework of precision holography. In this regard, 11d supergravity may provide a more favorable starting point, as it allows one to extract universal logarithmic contributions from a supergravity one-loop analysis \cite{Bhattacharyya:2012ye}, potentially extending beyond the simplest AdS vacuum as discussed in \cite{Hristov:2021zai,Bobev:2023dwx}, while consistently neglecting stringy massive modes that are not captured within the supergravity action.

\section*{Acknowledgments}

We are grateful to Nikolay Bobev, Gamseung Han, Dane Jeon, Suhyun Lee, Valentin Reys, Ashoke Sen, and Saman Soltani for useful discussions. This research is supported in part by the National Research Foundation of Korea (NRF) grant funded by the Korean government (MSIT), Grant No. RS-2024-00449284; by the Sogang University Research Grant No. 202410008.01; and by the Basic Science Research Program of the NRF funded by the Ministry of Education through the Center for Quantum Spacetime (CQUeST), Grant No. RS-2020-NR049598.

\appendix

\section{Conventions}\label{app:convention}

\subsection{Differential forms}\label{app:convention:diff}
In a $D$-dimensional spacetime, the Hodge star operator acts on a $p$-form as
\begin{equation}
	*(dx^{\mu_1}\wedge\cdots\wedge dx^{\mu_p})=\frac{1}{(D-p)!}\epsilon^{\mu_1\cdots \mu_p}{}_{\nu_{1}\cdots \nu_{D-p}}dx^{\nu_{1}}\wedge\cdots\wedge dx^{\nu_{D-p}}\,.\label{Hodge:dual}
\end{equation}
The totally anti-symmetric tensor is defined in the coordinate basis as
\begin{equation}
	\epsilon_{\mu_1\cdots\mu_D}=\begin{cases}
		\sqrt{|g|} & (\text{even permutation}) \\
		-\sqrt{|g|} & (\text{odd permutation}) \\
		0 & (\text{otherwise})
	\end{cases}\,,\label{epsilon}
\end{equation}
where $g$ denotes the determinant of the $D$-dim metric. The Hodge star operator satisfies the identity
\begin{align}
    *(*\omega_p)=(-1)^{p(D-p)+s}\omega_p\,,
\end{align}
where $s = 0$ for Euclidean signature and $s = 1$ for Lorentzian signature respectively. Hence in 10d spacetime, we define (anti) self-duality of a 5-form differently depending on the signature as 
\begin{align}
    \text{Lorentzian}:\quad *_\text{L}F_5&=\begin{cases}
        F_5 & (\text{self-dual}) \\
        -F_5 & (\text{anti self-dual})
    \end{cases}\,,\\
    \text{Euclidean}:\quad *F_5&=\begin{cases}
        i F_5 & (\text{self-dual}) \\
        -i F_5 & (\text{anti self-dual})
    \end{cases}\,.
\end{align}
For notational convenience, we omit the subscript ``E'' in the Euclidean signature.

\medskip

We also define the interior product between a $p$-form $\omega^{(p)}$ and a vector field $v=v^\mu\partial_\mu$ as
\begin{equation}
	\ri_v\omega^{(p)}=\fft{1}{(p-1)!}v^\nu\omega_{\nu\mu_1\cdots\mu_{p-1}}dx^{\mu_1}\wedge\cdots\wedge dx^{\mu_{p-1}}\,.
\end{equation}
Note that this definition is valid in both Euclidean and Lorentzian signatures.

\medskip

Based on the above summarized conventions, below we provide useful identities. Here $A_p$ and $B_q$ denote differential forms of degree $p$ and $q$, $D$ is the spacetime dimension, $s=0$ for Euclidean signature and $s=1$ for Lorentzian signature.
\begin{subequations}
 \begin{align}
    d (A_p \wedge B_q)&=d A_p \wedge B_q + (-1)^p A_p \wedge d B_q \\
    \ri_v(A_p \wedge B_q)&=\ri_v A_p \wedge B_q+ (-1)^p A_p \wedge \ri_v B_q\\
    (*)^2 A_p&=(-1)^{p(D-p)+s}A_p\\
    A_p\wedge *B_p&=B_p\wedge*A_p\\
    \ri_v (*A_p)&= *((-1)^pv\wedge A_p)\label{Identity:3}\\
    *(\ri_v A_p)&= (-1)^{p-1} v\wedge * A_p\\
    *(v\wedge\ri_v*A_p)&=(-1)^{p(D-p)+s}((\ri_vv)A_p-v\wedge\ri_v A_p) \label{Identity:5}
\end{align}   
\end{subequations}
%

\subsection{Wick rotation}\label{app:convention:Wick}
Consider the Wick rotation from a Lorentzian manifold to a Euclidean manifold ($D$-dim),
\begin{align}
    X_\text{L}^0 = -i X^0\qquad\&\qquad X_{\text{L}\,0}=i X_0\,, \label{Wick}
\end{align}
where we did not change the order of coordinates by taking $X_\text{L}^0 = -i X^0$ instead of $X_\text{L}^0 = -i X^D$ following \cite{Bilal:2003es}. It is worth mentioning that contracted contravariant/covariant indices are not affected by the Wick rotation (\ref{Wick}). As a consequence, differential forms remain unchanged under the Wick rotation (\ref{Wick}). 

\medskip

One must be careful with the Hodge star operator (\ref{Hodge:dual}) under the Wick rotation (\ref{Wick}) though. It changes under the Wick rotation (\ref{Wick}) as
\begin{align}
    *_\text{L}\omega_p=-i* \omega_p\,,
\end{align}
because we adopt the following convention of the anti-symmetric tensor
\begin{align}
    \epsilon^\text{(E)}_{012\cdots D-1}\neq -i\sqrt{g}~~\big(\text{natural from (\ref{Wick}) and}~\epsilon^\text{(L)}_{012\cdots D-1}=\sqrt{-g_\text{L}}\big)\quad\text{but}\quad \epsilon^\text{(E)}_{012\cdots D-1}=\sqrt{g}
\end{align}
according to (\ref{epsilon}).

\section{Gauge Invariance of the PST-IIB Action}\label{app:gauge-inv}
In this Appendix, we examine the invariance of the PST-IIB action \eqref{PST-IIB:action} under the gauge transformations \eqref{PST-IIB:gg}. Since the gauge transformations of interest act non-trivially only on the fields $C_4$ and $a$, we restrict our attention to variations with respect to these fields.

\medskip

We begin by using the identity
\begin{align}
    \tF_5 \wedge * \tF_5 + \ri_v \mF_5 \wedge * \ri_v \mF_5 = -2i v \wedge \tF_5 \wedge \ri_v \mF_5
\end{align}
to rewrite the PST-IIB action \eqref{PST-IIB:action}, where $\mF_5$ is defined in \eqref{calF5} and satisfies the anti-self-duality relation $*\mF_5 = -i \mF_5$ by construction. The general variation of the action then reads
\begin{equation}
\begin{split}
    \delta S_\text{PST}&=\frac{1}{8\kappa^2}\delta\bigg( \int_\mM\tF_5\wedge*\tF_5+\ri_v\mF_5\wedge *\ri_v\mF_5+2i C_4\wedge H_3\wedge F_3\bigg)
    \\&=-\frac{i}{4\kappa^2}\int_\mM\bigg( \tF_5\wedge\delta\tF_5+2v\wedge\delta\tF_5\wedge\ri_v\mF_5-\delta C_4\wedge H_3\wedge F_3\bigg)
    \\&\quad-\frac{i}{4\kappa^2}\int_{\mM}\bigg(-\frac{1}{\sqrt{-|\partial a|^2}}d\delta a\wedge v\wedge\ri_v\mF_5\wedge\ri_v\mF_5\bigg)
    \\&=-\frac{i}{4\kappa^2}\int_\mM\bigg(2\delta C_4\wedge d(v\wedge\ri_v\mF_5)+\delta a\,d(\frac{1}{\sqrt{-|\partial a|^2}}v\wedge\ri_v\mF_5\wedge\ri_v\mF_5)
    \bigg)
    \\&\quad +\frac{i}{4\kappa^2}\int_{\partial\mM}\bigg(\delta C_4\wedge \tF_5+2\delta C_4\wedge  v\wedge\ri_v\mF_5+\frac{1}{\sqrt{-|\partial a|^2}}\delta a\wedge v\wedge\ri_v\mF_5\wedge\ri_v\mF_5
    \bigg)\,.
\end{split}
\label{eq:general_variation}
\end{equation}
In the last step, we integrated by parts and used the Bianchi identity $d\tF_5 = H_3 \wedge F_3$. 

\medskip

Upon substituting the explicit gauge transformations defined in \eqref{PST-IIB:gg} into the general variation \eqref{eq:general_variation}, the bulk terms cancel identically, and the variation reduces to a pure boundary term. Accordingly, the analysis of gauge invariance proceeds by examining these remaining boundary integrals for each symmetry. 
\begin{itemize}
    \item For the $\delta_\eta$ transformation defined by $\delta_\eta a = \eta$ and $\delta_\eta C_4 = - \frac{\eta}{\sqrt{-|\partial a|^2}} \ri_v \mathcal{F}_5$, the  variation reduces to:
    \begin{equation}
        \delta_\eta S_\text{PST}  =- \frac{i}{4\kappa^2} \int_{\partial \mM} \frac{\eta}{\sqrt{-|\partial a|^2}} \left( \tF_5 + v \wedge \ri_v \mF_5 \right) \wedge \ri_v \mF_5 \,.
    \end{equation}
    This contribution vanishes provided that we impose the boundary condition $\eta|_{\partial \mM} = 0\,.$
    
    \item  For the $\delta_\ell$ transformation defined by $\delta_\ell a = 0$ and $\delta_\ell C_4 = \ell_4$ with $da \wedge d\ell_4 = 0$, the variation reduces to:
    \begin{equation}
        \delta_\ell S_\text{PST} = \frac{i}{4\kappa^2} \int_{\partial \mM} \tF_5 \wedge \ell_4 \,.
    \end{equation}
    This contribution vanishes provided that we impose the boundary condition $\ell_4|_{\partial \mM} = 0\,.$
\end{itemize}
In conclusion, the PST-IIB action \eqref{PST-IIB:action} is gauge invariant provided that the gauge parameters vanish on the boundary.

\bibliographystyle{JHEP}
\bibliography{PSTOnShell}

\end{document}